\documentclass[twocolumn]{aastex631}
\usepackage{graphicx}
\usepackage{bm}
\usepackage{physics}

\usepackage{CJKutf8}            

\begin{document}
\title{Spectroscopy-informed Deep Learning Searches for Photometric Little Red Dots in JWST Surveys} 
\author[0009-0006-2619-0107]{Chenxuan Zhang \begin{CJK*}{UTF8}{gkai} (张辰轩) \end{CJK*}}
\affiliation{Department of Astronomy, School of Physics, Huazhong University of Science and Technology, Luoyu Road 1037, Wuhan, China}
\affiliation{Hubei Fundamental Research Center for Physics, Wuhan, Hubei 430074, China}
\affiliation{Hubei Key Laboratory of Gravitation and Quantum Physics, Huazhong University of Science and Technology, Wuhan, Hubei 430074, China}

\author[0000-0002-0123-9246]{Huanian Zhang* \begin{CJK*}{UTF8}{gkai} (张华年) \end{CJK*}}
\affiliation{Department of Astronomy, School of Physics, Huazhong University of Science and Technology, Luoyu Road 1037, Wuhan, China}
\affiliation{Hubei Fundamental Research Center for Physics, Wuhan, Hubei 430074, China}
\affiliation{Hubei Key Laboratory of Gravitation and Quantum Physics, Huazhong University of Science and Technology, Wuhan, Hubei 430074, China}
\correspondingauthor{Huanian Zhang}
\email{huanian@hust.edu.cn}

\author[0000-0003-4773-4987]{Qingwen Wu* \begin{CJK*}{UTF8}{gkai} (吴庆文) \end{CJK*}}
\affiliation{Department of Astronomy, School of Physics, Huazhong University of Science and Technology, Luoyu Road 1037, Wuhan, China}
\affiliation{Hubei Fundamental Research Center for Physics, Wuhan, Hubei 430074, China}
\affiliation{Hubei Key Laboratory of Gravitation and Quantum Physics, Huazhong University of Science and Technology, Wuhan, Hubei 430074, China}
\correspondingauthor{Qingwen Wu}
\email{qwwu@hust.edu.cn}

\begin{abstract}

Little Red Dots (LRDs) are a population of compact, red sources discovered at high redshift with JWST that exhibit distinctive V-shaped spectral energy distributions (SEDs). Existing LRD samples are typically identified using a variety of color-based selection criteria, which can lead to differences in the populations recovered by different studies. We develop a cross-modal machine-learning framework that combines matched NIRCam imaging and NIRSpec/PRISM spectroscopy to transfer spectroscopic information into imaging-based representations, enabling LRD identification from JWST/NIRCam imaging alone. Applied to spectroscopically observed high-redshift galaxies previously classified as non-LRDs, our model identifies 35 new LRD candidates. The majority satisfy at least one independent LRD color selection criterion, and among the 13 candidates with medium- or high-resolution NIRSpec grating spectra, 10 exhibit broad H$\alpha$ emission—a hallmark of LRDs. We further apply the classifier to public NIRCam imaging from the Dawn JWST Archive and two independent JWST surveys, yielding 357 and 23 additional LRD candidates, respectively. The candidates recovered by our method span a broader range of observed SED properties than those in existing samples, suggesting that current color-selected samples may not fully capture the diversity of LRD-like systems present in JWST surveys. Our findings highlight the value of spectroscopy-informed photometric classification as a powerful strategy for systematically mining large-area JWST imaging datasets and provide a practical framework for future studies of this emerging population.

\end{abstract}


\section{Introduction}  \label{sec:intro}

The James Webb Space Telescope (JWST) has revealed a previously unknown population of compact red sources at high redshift known as little red dots (LRDs; \citealt{Matthee2024}). LRDs are characterized by distinctive V-shaped spectral energy distributions (SEDs), with blue rest-frame ultraviolet continua and much redder optical-to-near-infrared emission. They are also extremely compact and often exhibit broad Balmer emission lines \citep{Greene2024, Matthee2024, Hviding2025}. Since their discovery, hundreds of LRDs have been reported across multiple JWST surveys spanning redshifts $z\sim3$--10 \citep{Kokorev2024, Barro2024, Kocevski2025, Labbe2025, Hviding2025, deGraaff2025, Barro2026}. Their unexpectedly large abundance has attracted considerable attention: measured comoving number densities reach $\Phi \sim 10^{-5}$--$10^{-4},{\rm cMpc^{-3}}$ at $z>4$ \citep{Kokorev2024, Matthee2024, Kocevski2025, Akins2025, Rinaldi2026}. Such high number densities are difficult to reconcile with current models of early black-hole and galaxy growth, making LRDs one of the most intriguing populations discovered by JWST.

The physical nature of LRDs remains actively debated. Their compact morphologies, broad Balmer emission lines with exponential profiles, V-shaped SEDs, and prominent Balmer breaks are widely interpreted as signatures of accreting supermassive black holes embedded within dense gaseous environments, where the dense gas reprocess the intrinsic nuclear emission into the observed red continuum \citep{Inayoshi&Maiolino2025, Inayoshi&Luis2025, Naidu2025, Pacucci2025, Rusakov2026, Sneppen2026, Chenchanghao2026, Sunwendy2026, Begelman2026}.  About half of the LRD population shows no evident absorption features in their broad lines and a weak Balmer break, suggesting that the dense gas surrounding the black holes is anisotropic \citep[e.g.,][]{Davis2026, Matthee2026}. Furthermore, recent observations have identified a population of little blue dots (LBDs) that share many similarities with LRDs—including broad emission lines and compact morphologies—but exhibit blue rest-frame UV continua instead of red ones \citep[e.g.,][]{Brazzini2025, Brazzini2026}. This dichotomy suggests that orientation effects may play an important role in shaping the observed properties of these sources. \citet{Zhangchenxuan2025} proposed that the infrared emission of LRDs may originate from the dense gas in outer gravitationally unstable regions of the accretion disk \citep{Chenyixian2026, Zwick2026}, and the absorbing gas could be supplied by the accretion flow itself or associated winds. Beyond possible special nuclear physics, the host galaxies and large-scale environments of LRDs may also be distinctive. \citet{Pacucci2025} proposed that low-spin dark matter halos can naturally explain the abundance, compactness, and redshift evolution of the observed LRD population. \cite{ZhangC2026b} found that LRDs predominantly reside in under-dense regions and their dark matter halo masses grow rapidly from $\sim10^{10} M_{\odot}$ at $z\sim7$ to $\sim10^{11.3} M_{\odot}$ at $z\sim3$, which roughly approach the halo masses of typical galaxies at lower redshifts though still substantially lower than that of luminous quasars estimated from wide-field clustering analyses \citep{Meng2026}. This environmental and halo-mass evolution may provide a key link between the small-scale black-hole accretion physics and the large-scale structure formation, yet the detailed connection remains unclear.

Despite the rapidly increasing number of known LRDs, their classification remains strongly dependent on the adopted selection strategy, with color–color cuts being widely employed \citep[e.g.,][]{Kokorev2024, Akins2025, Barro2026, Rinaldi2026}. It should be noted that different thresholds will evidently affect sample completeness. Strict color selections can achieve high purity but preferentially identify the reddest and often optically brightest systems, potentially missing less extreme members of the population \citep[e.g.,][]{Rinaldi2026}. Spectroscopic identification provides the most direct approach since the characteristic V-shaped continuum and broad emission-line features of LRDs can be identified from their rest-frame optical spectra \citep{Hviding2025, deGraaff2025}. However, spectroscopic samples remain limited in size and are themselves subject to selection effects arising from target selection and observational strategies, which may bias the inferred properties of the spectroscopic LRD population \citep{Ginolfi2026}.

The diversity of LRD SEDs, combined with the strong selection dependence of existing samples, motivates the development of new approaches capable of recovering a more complete and representative population of LRDs. Machine learning has become an essential tool in observational astronomy \citep[e.g.,][]{Dieleman2015, Pasquet2019, Walmsley2022, Ye2024, ZhangHN2025}. Recent studies have shown its potential for LRD identification; \cite{Ginolfi2026} demonstrated that spectroscopically confirmed LRDs occupy coherent regions in a multidimensional photometric feature space, enabling the discovery of additional candidates without predefined color cuts. Existing data-driven approaches to LRD identification have primarily relied on catalog-level photometric and morphological measurements. While such methods have proven effective for candidate selection, they necessarily compress each source into a small set of summary quantities and do not fully exploit the information content of JWST observations. Pixel-level NIRCam images retain spatial structure, morphology, and surface-brightness distributions, while NIRSpec spectra encode wavelength-resolved continuum shapes and emission-line properties that are central to the identification of LRDs. Features such as the characteristic V-shaped continuum, Balmer break, and broad H$\alpha$ emission are only partially captured by broadband photometry and may be diluted or lost when each filter is represented by a single integrated flux \citep{Merz2026}. As a result, the complementary information contained in imaging and spectroscopy remains largely underutilized in current LRD searches.

In this work, we develop a cross-modal learning framework that leverages both JWST/NIRCam imaging and NIRSpec/PRISM spectroscopy during training. By aligning image and spectral representations through a contrastive-learning objective, the model transfers spectroscopic information into the image embedding space, encouraging the image encoder to learn features associated with the physical characteristics of LRDs. Once trained, the classifier requires only NIRCam imaging in up to seven bands and can therefore be applied efficiently to the much larger population of sources lacking spectroscopy or reliable photometric redshifts. This approach combines the diagnostic power of spectroscopy with the scalability of imaging surveys, providing a new pathway for constructing more complete and homogeneous samples of LRD candidates. 
We then apply the classifier to NIRCam imaging data from the Dawn JWST Archive \citep[DJA,][]{Heintz2024, deGraaff2025DJA} to construct a consistently selected catalog of new LRD candidates across a wide range of JWST fields. Finally, we apply the same classification model to imaging data from two independent surveys not used during training—the Slitless Areal Pure-Parallel High-Redshift Emission Survey (SAPPHIRES; \citealt{Sun2025}) and the Bias-free Extragalactic Analysis for Cosmic Origins with NIRCam survey (BEACON; \citealt{Morishita2025})—to identify LRD candidates in these new survey environments, thereby illustrating the applicability of the selection pipeline to additional surveys beyond the original training sample.

The paper is organized as follows. In Sections~\ref{sec:catalog} and \ref{sec:data}, we describe the training sample and the observational data used in this work. In Section~\ref{sec:model}, we present the network architecture and training procedure. In Section~\ref{sec:result}, we evaluate the classifier performance and present the resulting LRD candidate samples. In Section \ref{sec:discussion}, we present the various properties of the newly identified LRDs and discuss the possible limitations. Finally, in Section~\ref{sec:conclusion}, we summarize our main results.

\section{Training Catalog} \label{sec:catalog}

\subsection{Little Red Dots} \label{subsec:lrd}

We construct the LRD sample from compilations of JWST/NIRSpec-confirmed LRDs reported by \citet{deGraaff2025}, \citet{Hviding2025}, and \citet{Zhangchenxuan2025}. Although the selection criteria are not the same in these studies, all sources have spectroscopic evidence supporting their LRD classification. We visually inspect the spectra/images for the above LRDs, and reclassify eight ambiguous sources with extended morphologies or SEDs inconsistent with our adopted LRD definition. After requiring matched NIRCam imaging, a PRISM spectrum, and at least four valid filters, the final positive sample contains 127 spectroscopically confirmed LRDs.

To reduce potential contamination in the negative training sample, we compile a masking catalog of photometrically selected LRD candidates from \citet{Kocevski2025}, \citet{Kokorev2024}, \citet{Barro2024}, \citet{PerezGonzalez2024}, \citet{Labbe2025}, \citet{Zhangzijian2026}, and \citet{Barro2026}. Because these candidates lack spectroscopic confirmation and are selected using heterogeneous criteria, they are not used as positive training examples. Instead, we exclude these sources from the high-redshift galaxy sample in Section~\ref{subsec:hz} used to construct negative examples, which can reduce the risk that genuine LRDs are mistakenly treated as negative samples.

\subsection{Non-LRD High-redshift Galaxy} \label{subsec:hz}

We construct the non-LRD high-redshift galaxy sample from the DJA spectroscopic redshift catalog \citep{Heintz2024, deGraaff2025DJA}. Sources are selected to satisfy $z_{\rm spec}>3$, a spectroscopic quality flag of 3, and NIRSpec PRISM/CLEAR observations. This selection yields a comparison sample of high-redshift galaxies spanning a broad range of redshifts and galaxy properties.

To minimize contamination from potential LRDs, we remove the LRD candidate catalogs as described in Section~\ref{subsec:lrd}. After removing duplicate observations and requiring sufficient NIRCam coverage ($N_{\rm filter} \geq 4$), the final non-LRD high-redshift galaxy sample contains 6,702 spectroscopically confirmed sources and is used as the negative class. The sample has a median redshift of $z=4.3$, with more than 90\% of the sources at $z=3.0-7.5$. Its median F444W magnitude is 27.4 AB, with more than 90\% spanning 24.5--29.5 AB mag.

\subsection{High-redshift Quiescent Galaxy} \label{subsec:qg}

High-redshift quiescent and recently dormant galaxies constitute an important hard-negative population, as their compact morphologies and prominent Balmer or 4000~\AA\ breaks can mimic the red optical continua characteristic of LRDs. To assemble this subset, we draw from published high-redshift quiescent galaxy samples \citep{Ito2026, Baker2025, Zhangyunchong2026}, and supplement these by screening the remaining PRISM comparison spectra for objects exhibiting a significant Balmer break alongside weak H$\alpha$ emission, following the empirical diagnostics of \citet{CoveloPaz2026}. This process yields a total of 188 sources. During the training process as described below, we assign the quiescent galaxy sample a weight three times higher than that of the normal negative sample. This weighting scheme is motivated by the comparatively small size of the quiescent sample and the strong similarity of its red optical continuum SED to that of LRDs, which makes it a critically informative hard-negative population that warrants enhanced emphasis during model training. 

\subsection{Brown Dwarf} \label{subsec:bd}

Brown dwarfs (BDs) are a well-known contaminant of color-selected LRD samples because their near-infrared SEDs can mimic red-color criteria of LRDs \citep{Langeroodi2023}. We therefore construct the BD sample using two complementary components: spectroscopically confirmed objects observed with JWST and forward-modeled synthetic observations.

The observational component consists of spectroscopically confirmed BDs compiled from \citet{Langeroodi2023, Tuzhijun2025a, Tuzhijun2025b}. By cross-matching these catalogs with publicly available NIRSpec/PRISM observations in the DJA, we identify 14 sources with usable spectra. Because the number of observed JWST BDs remains limited, we augment the sample with 700 synthetic observations generated from BT-Settl stellar-atmosphere models \citep{Allard2012}. The simulations cover effective temperatures of $500$--$2400\ \mathrm{K}$, surface gravities of $\log g=4.5$--$5.5$, and solar metallicity.

We generate the synthetic observations by forward modeling the BT-Settl spectra through the JWST/NIRSpec/PRISM and NIRCam response functions. To accurately reproduce the spatial characteristics of the data, the spectra are used to generate images that include a convolution with a point spread function (PSF). Finally, the simulation is completed by adding a realistic background emission, mirroring the sky and instrumental backgrounds that will be presented in the science frames, ensuring the synthetic data are a robust testbed for data analysis and interpretation.

\section{Data} \label{sec:data}

\subsection{NIRCam Imaging Data} \label{subsec:image}

We use multi-band JWST/NIRCam imaging cutouts retrieved from the Dawn JWST Archive (DJA) with the \texttt{grizli} pipeline. To provide a uniform input representation for the image encoder, we extract fixed-size $32\times32$ pixel cutouts for all sources. At the adopted pixel scale of $0\farcs05\,\mathrm{pixel}^{-1}$, each cutout corresponds to a field of view of $1\farcs6\times1\farcs6$ centered on the catalog position of the source.

The image input consists of seven NIRCam filters: F115W, F150W, and F200W from the short-wavelength channels, and F277W, F356W, F410M, and F444W from the long-wavelength channels. These bands provide continuous wavelength coverage from approximately $1$--$5\,\mu$m and are among the most commonly available NIRCam observations in extragalactic JWST programs. For filters lacking available observations, the corresponding channels are zero-padded. Sources with fewer than four valid NIRCam bands are excluded from the analysis.


\subsection{NIRSpec Spectroscopic Data} \label{subsec:Spectroscopic}

The spectra are retrieved from the Dawn JWST Archive in the \texttt{grizli}-reduced \textsc{Spec1D} format \citep{Heintz2024DJA, deGraaff2025DJA}. The NIRSpec PRISM/CLEAR observations provide continuous wavelength coverage from $0.54$ to $5.52,\mu$m, enabling simultaneous measurements of rest-frame ultraviolet and optical continuum features over a broad redshift range. This wavelength coverage encompasses many of the key spectral signatures associated with LRDs, including the characteristic V-shaped continuum, Balmer-break features, and prominent emission lines such as H$\alpha$.


\section{Deep-learning Model} \label{sec:model}

\begin{figure*}[!t]
    \centering
    \includegraphics[width=\textwidth]{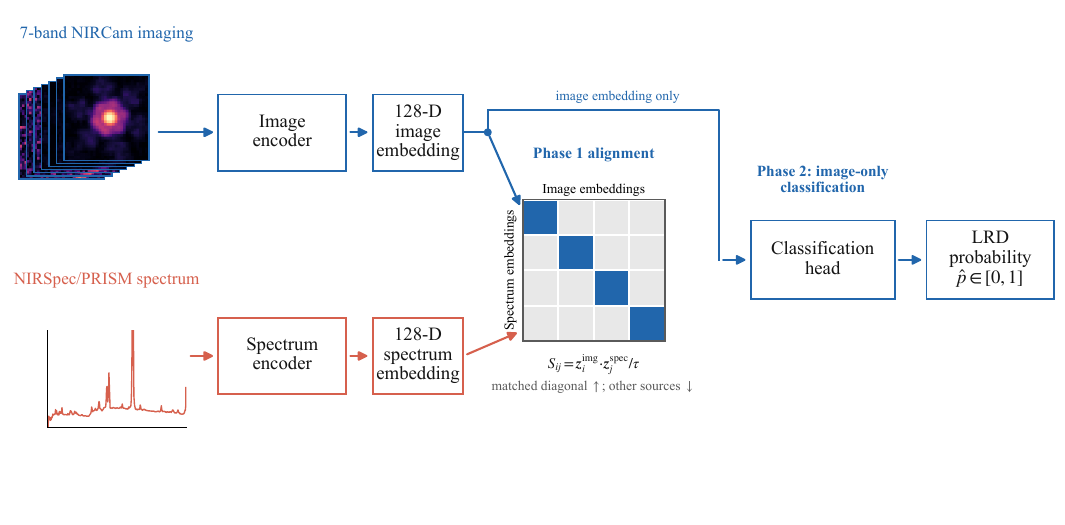}
    \caption{Cross-modal LRD classification framework. Phase~1 aligns paired NIRCam image and NIRSpec/PRISM spectrum embeddings in a shared space using a bidirectional InfoNCE objective. In Phase~2, the encoders are frozen, and only an image-based classification head is trained. Inference requires only NIRCam imaging.}
    \label{fig:architecture}
\end{figure*}

Identifying LRDs in large JWST imaging surveys is challenging because they are rare, and their most diagnostic features—including broad Balmer emission lines and the V-shaped UV-to-optical SED—are encoded primarily in spectroscopic data. In our spectroscopic training sample, LRDs account for only $1.8\%$ of sources, creating a severe class imbalance for standard classifiers. Although a supervised convolutional neural network (CNN) trained on imaging alone can identify LRD candidates with reasonable performance, cross-modal contrastive learning provides a means of incorporating complementary information from images and spectra into learned representations. AstroCLIP \citep{Parker2024} demonstrates the potential of this approach, outperforming a state-of-the-art self-supervised image-only model in morphology classification, with higher recall, and achieving approximately twice the performance in photometric-redshift estimation and galaxy-property prediction, as measured by the coefficient of determination ($R^2$). These results highlight the potential of cross-modal learning to improve the utility of image representations for downstream tasks while preserving modality-specific information. Motivated by these findings, we develop a two-phase framework that first aligns paired NIRCam images and NIRSpec spectra through contrastive learning without using class labels, and then trains an LRD classifier on the learned image representations. This strategy incorporates spectroscopic information during training while requiring only NIRCam imaging at inference (Figure~\ref{fig:architecture}).


The framework comprises an image encoder $f_{\rm img}$, a spectrum encoder $f_{\rm spec}$, and a binary classification head $g$. During Phase~1, paired images and spectra are aligned via a cross-modal contrastive objective, whereby spectroscopic information is transferred into the image representation. During Phase~2, the encoders are frozen, and the classifier is trained on the resulting image embeddings. At inference time, $f_{\rm spec}$ is discarded, and the model predicts an LRD probability $p\in[0,1]$ from seven-band NIRCam imaging alone.

\subsection{Data Preprocessing} \label{subsec:preprocessing}

\textbf{Imaging.}
The seven NIRCam bands are normalized jointly for each source. We define a single normalization factor $q$ as the $99.5$th percentile of the absolute pixel values across all available bands and divide all channels by this common scale. This preserves the relative flux ratios between filters, which contain the broadband color information, while providing a consistent dynamic range across sources. The normalized images are then transformed using an inverse hyperbolic sine (\texttt{asinh}) stretch \citep{Lupton2004} with a softening parameter of $s=0.01$ to compress the dynamic range while retaining information from low-surface-brightness and negative background-subtracted pixels.

\textbf{Spectroscopy.}
The PRISM spectra are converted to $f_\lambda$. Three masked zero-value bins are appended so that the 476-pixel input is divisible by the adopted patch width. The valid flux pixels are normalized by their 99th percentile in absolute value. The final input has shape $(3,476)$ and contains normalized $f_\lambda$, clipped and scaled signal-to-noise ratio, and a binary validity mask.

\subsection{Network Architecture} \label{subsec:network}

The complete model consists of three components,
$\mathcal{M}=\{f_\mathrm{img}, f_\mathrm{spec}, g\}$: an image encoder, a spectrum encoder, and a classification head.

\textbf{Image Encoder $f_\mathrm{img}$.}
The image encoder maps a seven-band NIRCam cutout of size $(7,32,32)$ to a 128-dimensional unit-normalized embedding $\boldsymbol{z}^{\rm img}$. We adopt a widened CIFAR ResNet-20 architecture \citep{He2016} modified for seven-channel inputs. The network contains three residual stages, each consisting of three residual blocks, with channel widths of 32, 64, and 128, respectively. A two-layer nonlinear projection head maps the pooled feature vector to the final 128-dimensional embedding.

\textbf{Spectrum Encoder $f_\mathrm{spec}$.}
The spectrum encoder maps the preprocessed PRISM input of shape $(3,476)$ to a 128-dimensional unit-normalized embedding $\boldsymbol{z}^{\rm spec}$.
The three input channels correspond to the normalized $f_\lambda$ spectrum, the clipped and scaled signal-to-noise ratio, and the binary validity mask. We adopt a one-dimensional Vision Transformer-style architecture \citep{Dosovitskiy2020}, in which a non-overlapping one-dimensional convolution with kernel size and stride of seven divides the 476-pixel input into 68 spectral tokens with 128-dimensional embeddings.
A learnable \textsc{[cls]} token and positional embeddings are added before processing by three pre-LayerNorm Transformer encoder blocks \citep{Vaswani2017}.
The final \textsc{[cls]} representation is linearly projected into the shared 128-dimensional image--spectrum embedding space and $\ell_2$ normalized.

\textbf{Classification Head $g$.}
The classification head maps $\boldsymbol{z}^{\rm img}$ to an LRD logit via a two-layer MLP comprising a hidden linear layer, LayerNorm, ReLU activation, dropout, and a final linear output layer. The sigmoid transformation converts the output logit into the predicted LRD probability $p$.  

\subsection{Two-phase Training} \label{subsec:training}

The model is trained in two stages: cross-modal representation learning and supervised image classification. We adopted a stratified split with a target validation fraction of 20\% for both the LRD and negative classes. The resulting training and validation sets contain 6,079 and 1,462 sources, including 102 and 25 labeled LRDs, respectively. Stochastic data augmentation is applied only during training. NIRCam images undergo horizontal and vertical flips, right-angle and arbitrary rotations, and cutout masking outside the central region. We additionally apply multiplicative flux perturbations, Gaussian noise, and random channel dropout to improve robustness to photometric uncertainties and incomplete filter coverage. During Phase~1, spectra are augmented with uncertainty-scaled Gaussian noise, shifts of up to one native wavelength bin, and masking of 8-24 consecutive bins.

During Phase~1, the image and spectrum encoders are jointly optimized using a bidirectional InfoNCE loss \citep{vandenOord2018, Radford2021}, while the classification head remains frozen.
The temperature-scaled similarity between image and spectrum embeddings is

\begin{equation}
s_{ij}
=
\frac{
(\boldsymbol{z}^{\rm img}_{i})^{\mathsf T}
\boldsymbol{z}^{\rm spec}_{j}
}{\tau},
\qquad
i,j\in\mathcal{B}_{\rm spec},
\end{equation}
where $\tau$ is the temperature parameter. We initialize $\tau=0.07$ and learn it during Phase~1, constraining it to $0.01\leq\tau\leq0.5$.
The bidirectional contrastive loss is
\begin{equation}
\mathcal{L}_{\rm align}
=
-\frac{1}{2M}
\sum_{i}
\left[
\log
\frac{e^{s_{ii}}}{\sum_j e^{s_{ij}}}
+
\log
\frac{e^{s_{ii}}}{\sum_j e^{s_{ji}}}
\right],
\end{equation}
where $s_{ij}$ is the temperature-scaled cosine similarity between the
image and spectrum embeddings of samples $i$ and $j$.

During Phase~2, both encoders are frozen, and only the classification head is optimized using the image embeddings.
Let $\ell_i$ denote the predicted classification logit, $p_i=\sigma(\ell_i)$ the corresponding probability, and $y_i\in\{0,1\}$ the target label.
The classification objective combines a weighted focal loss \citep{Lin2017} and a pairwise logistic ranking loss:

\begin{equation}
\mathcal{L}_{\rm focal}
=
-\frac{1}{|\mathcal{B}|}
\sum_{i\in\mathcal{B}}
\alpha_{t,i}(1-p_{t,i})^\gamma
\log p_{t,i}.
\end{equation}
Here $p_{t,i}$ denotes the probability assigned to the ground-truth class, and $\alpha_{t,i}=\alpha$ for LRDs and $1-\alpha$ for negative sources. We adopt $\gamma=2$, following the standard focal-loss setting of \citet{Lin2017}, and set $\alpha=0.8$ to assign greater weight to the rare LRD class.

To further improve the separation between LRDs and negative examples, we introduce a RankNet-style pairwise logistic ranking loss:

\begin{equation}
\mathcal{L}_{\rm rank}
=
-\frac{1}{|\mathcal{P}||\mathcal{N}|}
\sum_{i\in\mathcal{P}}
\sum_{j\in\mathcal{N}}
\log
\sigma(\ell_i-\ell_j),
\end{equation}
where $\mathcal{P}$ and $\mathcal{N}$ denote positive and negative samples within the mini-batch, respectively.
This term encourages LRD candidates to receive larger logits than negative examples. The Phase~2 objective is $\mathcal{L}_{\rm focal}+0.2\mathcal{L}_{\rm rank}$.
Both phases use the AdamW optimizer \citep{Loshchilov2019}. We use a learning rate of $10^{-4}$ in both training phases, a weight decay of 0.01, and a dropout rate of 0.1 in the Phase~2 classification head. In Phase~1, the learning rate is linearly warmed up from $10^{-5}$ to $10^{-4}$ over the first five epochs and then decayed to $10^{-7}$ using a cosine schedule. In Phase~2, we apply cosine decay from $10^{-4}$ to $10^{-7}$ without warm-up. Training uses a batch size of 128. Each phase is trained with early stopping.

\subsection{Inference} \label{subsec:inference}

During inference, the spectrum encoder is discarded, and the LRD probability is predicted using only the image encoder $f_\mathrm{img}$ and classification head $g$. This enables application of the classifier to imaging-only samples without requiring spectroscopic observations.

To improve robustness against image orientation, we apply deterministic test-time augmentation (TTA). Each source is evaluated using eight geometrical transformations, consisting of four rotations ($0^\circ$, $90^\circ$, $180^\circ$, and $270^\circ$) with and without horizontal flipping. The final prediction is obtained by averaging the probabilities from all transformed inputs:

\begin{equation}
p_{\rm TTA} =
\frac{1}{8}\sum_{k=1}^{8}
\sigma\!\left[g\!\left(f_\mathrm{img}
(\mathcal{T}_k(\mathbf{x}))\right)\right],
\end{equation}
where $\mathcal{T}_k$ denotes the $k$th geometric transformation. 

Sources with $p_{\rm TTA}\geq\theta$ are classified as LRD candidates, where we adopt $\theta=0.80$ as the primary selection threshold throughout this work.

\section{Results} \label{sec:result}

\subsection{Validation performance} \label{subsec:performance}

Taking the catalog-labeled LRDs as the positive class, we evaluate classifier performance using precision, recall, F1 score, false-positive rate (FPR), and the area under the receiver-operating-characteristic curve (AUC--ROC). Precision and recall are defined as $\mathrm{TP}/(\mathrm{TP}+\mathrm{FP})$ and $\mathrm{TP}/(\mathrm{TP}+\mathrm{FN})$, respectively, while the F1 score is their harmonic mean. The false-positive rate is given by $\mathrm{FP}/(\mathrm{FP}+\mathrm{TN})$. Here, TP, FP, FN, and TN denote the numbers of true positives, false positives, false negatives, and true negatives, respectively. The relative importance of these metrics depends on the scientific objective. In this work, our primary goal is to construct as complete a sample of LRD candidates as possible, thereby expanding beyond existing color-selected samples and enabling the discovery of potentially overlooked members of the population. We therefore place particular emphasis on recall, which quantifies the fraction of known LRDs recovered by the classifier. Other applications may prioritize precision, especially when follow-up resources are limited and a purer candidate sample is desired. In such cases, a higher classification threshold could be adopted to reduce contamination at the expense of completeness.

\begin{figure}
    \centering
    \includegraphics[width=\columnwidth]{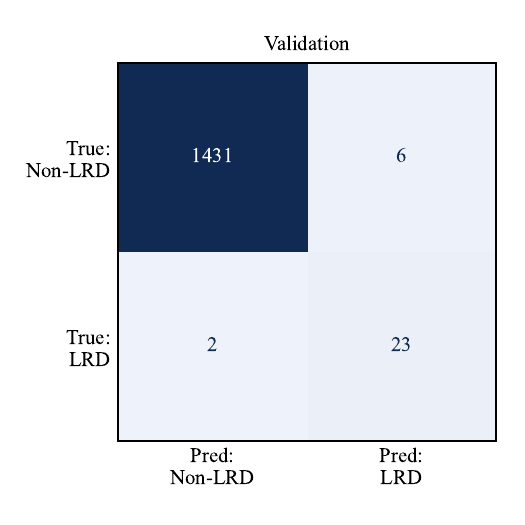}
    \caption{Validation-set confusion matrix at $\theta=0.80$. Rows and columns denote the true and predicted classes, respectively, and entries give the number of sources.}
    \label{fig:confusion}
\end{figure}

At the adopted classification threshold of $\theta = 0.80$, the eight-view test-time augmentation (TTA) model was evaluated on the internal model-selection validation set ($n = 1462$; 25 LRDs and 1,437 non-LRDs). The model correctly identifies 23 of the 25 labeled LRDs while classifying 6 labeled non-LRDs as LRDs. This corresponds to a recall of 0.92, a precision of 0.79, an F1 score of 0.85, and an AUC--ROC of 0.998. The resulting confusion matrix is shown in Figure~\ref{fig:confusion}.  The recall achieved by the cross-modal model is substantially higher than that of the imaging-only CNN (0.84). We adopt \(\theta=0.80\) as a compromise between precision and recall, prioritizing recall to construct a more complete sample of LRD candidates. We emphasize that the reported scores represent performance estimates on the internal validation set and should not be interpreted as calibrated posterior probabilities.


These performance metrics should be interpreted as estimates derived from the internal validation sample rather than precise measurements of the intrinsic completeness and purity of the broader LRD population. The validation set contains only 25 positive examples, meaning that the classification outcome of a single source changes the measured recall by 0.04. Consequently, statistical uncertainties remain significant, and a larger independent spectroscopic sample will be required to obtain more robust constraints on classifier performance.

Because the negative class is defined using existing catalogs and may contain uncataloged LRDs, the measured precision is likely conservative. All six validation-set false positives appear LRD-like upon visual inspection: sources RUBIES-UDS41-24447, GDS-BARRUFET-S156-4307, RUBIES-EGS51-13872, NEXUS-OBS3-33289, NEXUS-OBS5-11737, and MACS1149-5104172 (see those objects in Figures \ref{fig:fp_spectra_appendix_01} and \ref{fig:fp_spectra_appendix_02} in the Appendix). In other words, the reported precision should be interpreted as a lower bound, implying that the nominal precision may be conservative.

\subsection{Embedding Space} \label{subsec:embedding-space}

\begin{figure}
    \centering
    \includegraphics[width=\columnwidth]{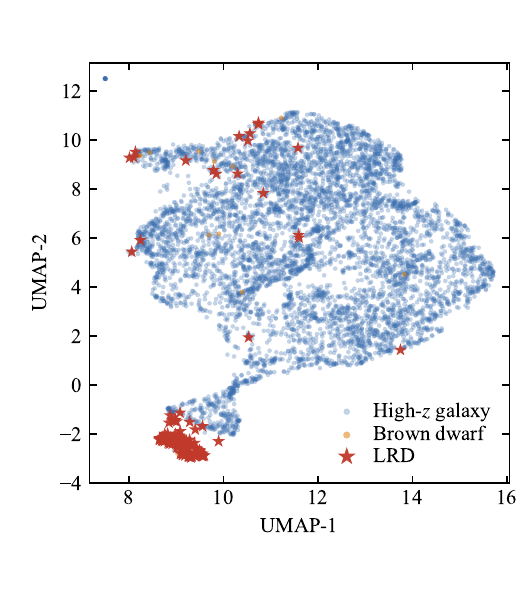}
    \caption{Two-dimensional UMAP projection of the 128-dimensional image embeddings for the labeled sample. Blue circles, orange circles, and red stars represent high-redshift galaxies, brown dwarfs, and LRDs, respectively. The displayed limits emphasize the LRD manifold.}
    \label{fig:umap}
\end{figure}

Figure~\ref{fig:umap} presents a two-dimensional UMAP projection \citep{McInnes2018} of the 128-dimensional image embeddings for the labeled sample. LRDs occupy a compact region in the lower-left portion of the embedding space and are largely separated from the broader high-redshift galaxy population, although some overlap remains between the two classes. The displayed limits are chosen to focus on this shared galaxy manifold; the detached branch containing the majority of the simulated brown dwarfs lies outside the plotted region, while the observed brown dwarfs and a small number of simulated objects remain visible. The partial overlap between LRDs and high-redshift galaxies is qualitatively similar to the photometric-feature embedding reported by \citet{Ginolfi2026} and is consistent with a continuous distribution of observed SED properties rather than a perfectly distinct population.

The overlap likely reflects a combination of intrinsic diversity within the LRD population, measurement uncertainties, and incompleteness in the current catalogue labels. In particular, several high-probability sources located in this boundary region satisfy independent LRD selection criteria (Section~\ref{subsec:spec-lrd}), suggesting that at least part of the overlap may arise from previously unrecognized LRDs within the sources labeled as ordinary high-redshift galaxies. We emphasize that UMAP is used here solely as a visualization tool to illustrate the structure of the learned embedding space; it does not define the classifier decision boundary or enter directly into the classification procedure.

\subsection{Spectroscopic LRD Candidates} \label{subsec:spec-lrd}

Applying the classifier to the full spectroscopic high-redshift galaxy sample yields 35 candidates at $p \geq 0.80$ that were previously labeled as non-LRDs. Six of the 35 candidates belong to the validation set and correspond to the six nominal false positives reported in Section~\ref{sec:result}. The NIRSpec/PRISM spectra of all 35 candidates are shown in Figures~\ref{fig:fp_spectra_appendix_01} and \ref{fig:fp_spectra_appendix_02} in the Appendix. Because spectroscopic observations are available for every candidate, their continuum and emission-line properties can be examined directly, providing an opportunity to assess whether the classifier is recovering genuine LRD candidates that are absent from existing catalogs rather than simply selecting contaminants.

\begin{figure*}[!ht]
    \centering
    \includegraphics[width=0.96\linewidth]{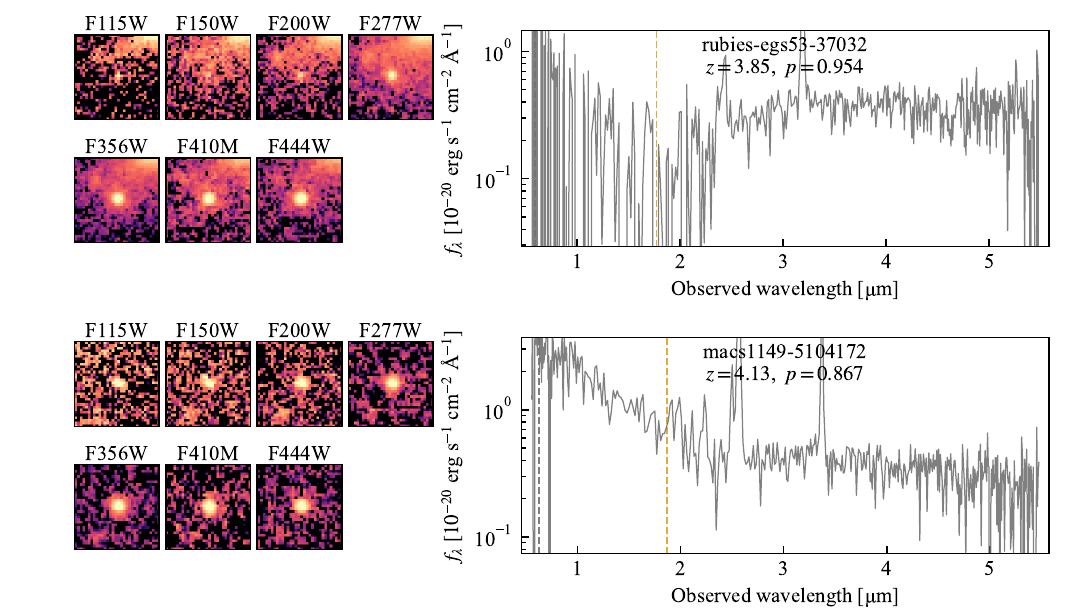}
    \caption{Two representative spectroscopic LRD candidates. The panels show the seven NIRCam cutouts and the NIRSpec/PRISM spectrum for each source. The black and orange dashed lines mark the expected Ly$\alpha$ and Balmer-break wavelengths, respectively. }
    \label{fig:candidate_showcase}
\end{figure*}

Figure~\ref{fig:candidate_showcase} presents two representative high-probability candidates. Both sources exhibit compact, red NIRCam morphologies and pronounced V-shaped continua in their PRISM spectra, closely resembling the spectral characteristics of spectroscopically confirmed LRDs. These examples demonstrate that the classifier is not simply identifying compact red sources, but is recovering objects that exhibit the key spectroscopic signatures of the LRD population. The consistency between the image-based predictions and the independently measured spectral properties suggests that the learned representations capture physically relevant information and underscores the power of the model to identify LRD candidates beyond those currently recognized in spectroscopic catalogs.

We define the Balmer-break strength as $B_{\rm Balmer}\equiv \widetilde{f}_{\nu,\rm red}/\widetilde{f}_{\nu,\rm blue}$, where $\widetilde{f}_{\nu,\rm red}$ and $\widetilde{f}_{\nu,\rm blue}$ are the median spectral flux densities in the rest-frame 4100--4300 and 3400--3600~\AA\ windows, respectively. Larger values indicate stronger Balmer breaks. Table~\ref{tab:fp_recovery} summarizes the positions, spectroscopic redshifts, classifier probabilities, F444W magnitudes, continuum slopes, Balmer-break strengths, and F444W compactness measurements of the 35 spectroscopic candidates. The sample spans a broad range of redshifts and continuum properties, with substantial variation in UV slopes, optical slopes, and Balmer-break strengths. This diversity indicates that the selected sources do not form a homogeneous population and may include objects with different physical origins or evolutionary states. At the same time, the prevalence of LRD-like spectral characteristics among many of the candidates suggests that a substantial fraction of them may represent previously unrecognized members of the LRD population. We therefore regard these sources as a candidate catalog for future investigation rather than a uniformly confirmed LRD sample. Targeted spectroscopic analyses and follow-up observations will be required to establish their nature and quantify the fraction of genuine LRDs recovered by the classifier.

\begin{deluxetable*}{rlccccccccc}
\tabletypesize{\scriptsize}
\tablewidth{0pt}
\tablecaption{Observed properties of the model-selected spectroscopic LRD candidates. \label{tab:fp_recovery}}
\tablehead{
\colhead{\#} & \colhead{ID} & \colhead{RA} & \colhead{Dec} & \colhead{$z_{\rm spec}$} & \colhead{$p_{\rm LRD}$} & \colhead{$m_{\rm F444W}$} & \colhead{$\beta_{\rm UV}$} & \colhead{$\beta_{\rm opt}$} & \colhead{$B_{\rm Balmer}$} & \colhead{$C_{\rm F444W}$}
}
\startdata
1 & rubies-uds23-166271 & 34.35326 & -5.11227 & 5.08 & 0.964 & 25.76 & $\cdots$ & $\cdots$ & $\cdots$ & 1.09 \\
2 & nexus-obs5-24361 & 268.48484 & 65.20720 & 3.56 & 0.961 & 24.23 & -0.56 & 1.34 & 4.81 & 1.11 \\
3 & rubies-uds41-41571 & 34.40709 & -5.25167 & 5.49 & 0.960 & 25.80 & -1.32 & -1.94 & 4.83 & 1.12 \\
4 & rubies-egs53-37032 & 214.84939 & 52.81182 & 3.85 & 0.954 & 25.76 & -1.83 & 2.46 & 0.92 & 1.10 \\
5 & macs0416-3105725 & 64.07687 & -24.10346 & 7.47 & 0.950 & 27.23 & -1.80 & 1.86 & 1.42 & 1.12 \\
6 & macs0417-202573 & 64.38110 & -11.83579 & 3.10 & 0.947 & 25.11 & -2.56 & 1.66 & 3.32 & 1.10 \\
7 & rubies-uds41-24447 & 34.38936 & -5.27909 & 5.26 & 0.947 & 25.75 & -2.85 & $\cdots$ & $\cdots$ & 1.26 \\
8 & rubies-uds31-977881 & 34.42962 & -5.11230 & 7.27 & 0.943 & 25.12 & -0.30 & -2.35 & 2.18 & 1.18 \\
9 & gds-barrufet-s156-4307 & 53.14063 & -27.88138 & 5.52 & 0.943 & 25.32 & -0.84 & 0.06 & 3.56 & 1.10 \\
10 & rubies-egs51-13872 & 215.13293 & 52.97071 & 5.26 & 0.942 & 26.36 & -2.48 & -0.74 & 1.29 & 1.11 \\
11 & nexus-obs5-3147 & 268.44255 & 65.14432 & 3.94 & 0.942 & 24.43 & -0.97 & -0.31 & 1.40 & 1.11 \\
12 & nexus-obs3-33289 & 268.40404 & 65.21211 & 4.05 & 0.935 & 24.90 & -1.34 & 1.22 & 1.31 & 1.11 \\
13 & abell370-2104488 & 39.98308 & -1.62414 & 4.05 & 0.934 & 27.60 & -2.82 & 1.95 & $\cdots$ & 1.12 \\
14 & jades-gds03-172975 & 53.08773 & -27.87124 & 4.74 & 0.929 & 27.21 & 0.42 & -0.07 & 2.08 & 1.11 \\
15 & macs0416-3108879 & 64.05249 & -24.08145 & 5.34 & 0.927 & 25.55 & $\cdots$ & 1.02 & $\cdots$ & 1.11 \\
16 & nexus-obs5-16223 & 268.39460 & 65.25418 & 3.88 & 0.920 & 25.12 & -1.55 & -0.59 & 1.03 & 1.12 \\
17 & rubies-uds21-176292 & 34.22810 & -5.09814 & 3.13 & 0.904 & 25.70 & -3.69 & 1.98 & 4.58 & 1.14 \\
18 & macs1149-5102050 & 177.38686 & 22.35191 & 5.20 & 0.897 & 27.23 & -0.76 & -0.06 & 0.92 & 1.14 \\
19 & gds-barrufet-s156-5607 & 53.12337 & -27.87054 & 3.59 & 0.897 & 26.19 & -1.81 & 1.06 & 4.01 & 1.12 \\
20 & rubies-uds1-19484 & 34.23243 & -5.28065 & 4.66 & 0.893 & 26.52 & -2.19 & 1.35 & 5.71 & 1.17 \\
21 & rubies-egs63-47179 & 214.89676 & 52.87580 & 6.13 & 0.892 & 27.64 & $\cdots$ & -0.38 & $\cdots$ & 1.14 \\
22 & nexus-obs5-11737 & 268.43292 & 65.19192 & 3.87 & 0.892 & 25.61 & -3.20 & 1.93 & 3.55 & 1.09 \\
23 & ceers-2355 & 215.00849 & 52.97797 & 6.12 & 0.884 & 25.83 & -0.84 & 0.39 & $\cdots$ & 1.10 \\
24 & capers-cos13-160883 & 150.07618 & 2.39483 & 7.66 & 0.874 & 27.17 & -0.24 & 0.26 & 0.64 & 1.09 \\
25 & capers-cos13-9924 & 150.11617 & 2.42746 & 4.00 & 0.869 & 27.21 & -1.90 & 0.37 & 3.98 & 1.01 \\
26 & macs1149-5104172 & 177.41265 & 22.36909 & 4.13 & 0.867 & 25.95 & -1.24 & -0.07 & 1.07 & 1.08 \\
27 & nexus-obs3-10347 & 268.52006 & 65.16487 & 5.75 & 0.866 & 24.21 & -1.58 & -0.64 & 1.00 & 1.11 \\
28 & nexus-obs3-4750 & 268.44622 & 65.13424 & 3.95 & 0.864 & 25.30 & $\cdots$ & 1.27 & $\cdots$ & 1.07 \\
29 & capers-cos13-9836 & 150.11319 & 2.42810 & 4.92 & 0.856 & 26.08 & -1.84 & 0.48 & 1.42 & 1.22 \\
30 & nexus-obs5-4316 & 268.35321 & 65.15233 & 6.03 & 0.853 & 25.24 & -1.11 & -0.90 & 1.22 & 1.12 \\
31 & rubies-uds3-42289 & 34.30365 & -5.24266 & 3.70 & 0.841 & 26.39 & -1.62 & 0.19 & 1.87 & 1.09 \\
32 & jades-gds-wide2-3044 & 53.12136 & -27.79479 & 4.12 & 0.833 & 28.20 & -3.15 & -2.26 & 2.21 & 1.22 \\
33 & rubies-uds42-826393 & 34.33032 & -5.28125 & 8.56 & 0.819 & 26.49 & -1.02 & 2.70 & 2.00 & 1.14 \\
34 & nexus-obs5-12185 & 268.45340 & 65.19404 & 7.40 & 0.813 & 26.39 & -1.25 & -0.44 & 1.58 & 1.06 \\
35 & rubies-uds22-114988 & 34.29794 & -5.18437 & 4.37 & 0.810 & 24.13 & $\cdots$ & 3.10 & $\cdots$ & 1.30 \\
\enddata
\tablecomments{$m_{\rm F444W}$ is the AB magnitude measured in a $0\farcs36$-diameter aperture. The continuum slopes are defined by $f_\lambda\propto\lambda^\beta$. $B_{\rm Balmer}$ is the Balmer-break strength, defined as the ratio of the median rest-frame $f_\nu$ at 4100--4300~\AA\ to that at 3400--3600~\AA. $C_{\rm F444W}=f_{0\farcs50}/f_{0\farcs36}$ is the F444W aperture-flux ratio. A value is listed as \nodata\ when either flux measurement is missing, non-finite, or non-positive. }
\end{deluxetable*}

\begin{figure*}[!ht]
    \centering
    \includegraphics[width=1\linewidth]{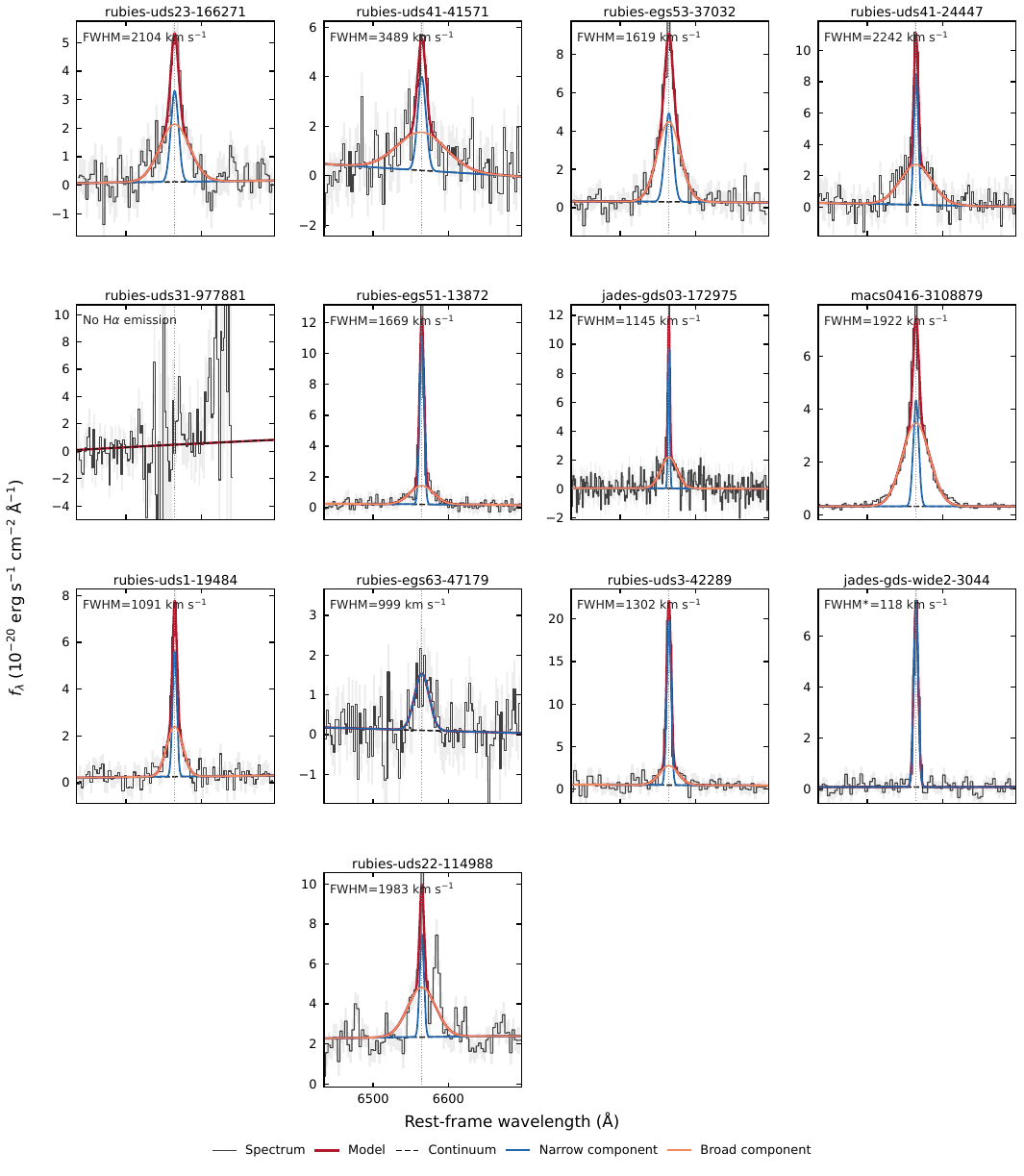}
    \caption{ H$\alpha$ fits for 13 candidates. Black line and gray shadow show the spectra and $1\sigma$ uncertainties. Red curves are the minimum-BIC models; dashed black curves are the continua, and blue and orange curves are the narrow and broad components added to the continuum. Labels give the intrinsic FWHM of the broad component for two-Gaussian models or of the line for single-Gaussian models; the asterisk marks the narrow-component width of jades-gds-wide2-3044.}
    \label{fig:Ha_emission_line}
\end{figure*}

Medium- or high-resolution NIRSpec grating spectra are available for 15 of the 35 spectroscopic candidates, with H$\alpha$ falling within the available wavelength coverage for 13 of these sources (Figure \ref{fig:Ha_emission_line}). To assess the presence of broad-line emission, we compare a continuum-only model with single- and two-Gaussian H$\alpha$ emission-line models using the Bayesian information criterion (BIC); the two-Gaussian model comprises a narrow and a broad component. The preferred model includes a broad H$\alpha$ component for 10 sources, a single narrow component for two sources, and no significant H$\alpha$ emission for one source (Table~\ref{tab:fp_grating_halpha}). For the candidates requiring a broad component, the inferred H$\alpha$ FWHM values span approximately $1100$--$3500,\mathrm{km\ s^{-1}}$, consistent with the broad Balmer emission commonly observed in spectroscopically confirmed LRDs. The high incidence of broad H$\alpha$ emission among these independently selected candidates provides direct spectroscopic support for their interpretation as members of the LRD population.

The three candidates without a BIC-preferred broad component illustrate the diversity of outcomes within the sample. The SED of rubies-uds31-977881 is visually more consistent with that of a quiescent galaxy. For this source, the continuum-only model is preferred, and the unresolved H$\alpha$ signal-to-noise ratio is only 0.9. While the absence of detectable H$\alpha$ emission does not by itself establish quiescence, it is consistent with this interpretation and highlights compact quiescent galaxies as a potential contaminant for image-based LRD selection. For jades-gds-wide2-3044, the measured intrinsic H$\alpha$ width of $118\pm45\mathrm{km\ s^{-1}}$ lies within the narrow-line regime reported for some spectroscopically identified LRDs \citep{Zhangzijian2026}. In contrast, rubies-egs63-47179 has a fitted H$\alpha$ FWHM of $999\pm260 \mathrm{km\ s^{-1}}$, close to the adopted upper limit for the narrow component. However, the line is detected with a modest signal-to-noise ratio of only 4.7, preventing a robust distinction between narrow- and broad-line interpretations. Higher signal-to-noise spectroscopy will therefore be required to determine the nature of this source and to characterize better the full diversity of emission-line properties among the candidate LRD population.

The prevalence of broad H$\alpha$ emission among the model-selected candidates provides an important test of the classifier. Because the candidate selection is performed using NIRCam imaging alone, the agreement with a key spectroscopic characteristic of many LRDs suggests that the model is not merely identifying compact red galaxies based on their observed colors or morphologies. Instead, the learned image representations appear to encode the underlying physical properties of the LRD population. More importantly, the grating spectroscopy was not used in the candidate selection process, so the emergence of broad-line systems within the selected sample demonstrates that the cross-modal training framework has successfully transferred spectroscopic information into the imaging representation. This consistency between image-based classifications and independent spectroscopic diagnostics supports the interpretation that many of the selected candidates are genuine LRDs and highlights the ability of the model to identify physically meaningful LRD-like systems beyond existing catalogs.

\begin{deluxetable*}{rlcccc}
\tabletypesize{\scriptsize}
\tablewidth{0pt}
\tablecaption{BIC-selected H$\alpha$ Models for the 13 Spectroscopic Candidates with Usable NIRSpec Grating Coverage. \label{tab:fp_grating_halpha}}
\tablehead{
\colhead{\#} & \colhead{ID} & \colhead{$z_{\rm spec}$} & \colhead{$\Delta\mathrm{BIC}$} & \colhead{Model} & \colhead{FWHM (km s$^{-1}$)}
}
\startdata
1 & rubies-uds23-166271 & 5.0758 & 10.9 & Broad & $2104 \pm 388$ \\
3 & rubies-uds41-41571 & 5.4828 & 6.5 & Broad & $3489 \pm 879$ \\
4 & rubies-egs53-37032 & 3.8515 & 7.4 & Broad & $1619 \pm 253$ \\
7 & rubies-uds41-24447 & 5.2571 & 41.7 & Broad & $2242 \pm 358$ \\
8 & rubies-uds31-977881 & 7.2818 & $-6.5*$ & No emission & $\cdots$ \\
10 & rubies-egs51-13872 & 5.2611 & 40.4 & Broad & $1669 \pm 300$ \\
14 & jades-gds03-172975 & 4.7417 & 54.7 & Broad & $1145 \pm 166$ \\
15 & macs0416-3108879 & 5.3340 & 1997.3 & Broad & $1922 \pm 51$ \\
20 & rubies-uds1-19484 & 4.6555 & 10.0 & Broad & $1091 \pm 238$ \\
21 & rubies-egs63-47179 & 6.1274 & -10.7 & Narrow & $999 \pm 260*$ \\
31 & rubies-uds3-42289 & 3.6990 & 6.9 & Broad & $1302 \pm 428$ \\
32 & jades-gds-wide2-3044 & 4.1157 & -6.2 & Narrow & $118 \pm 45*$ \\
35 & rubies-uds22-114988 & 4.3647 & 35.9 & Broad & $1983 \pm 349$ \\
\enddata
\tablecomments{Row numbers correspond to Table~\ref{tab:fp_recovery}. Positive $\Delta\mathrm{BIC}$ values favor the more complex model, while negative values favor the simpler model. The listed FWHM is the intrinsic width of the preferred component; for \texttt{rubies-uds31-977881}, the preferred model contains no emission line, indicated by the symbol \nodata.}
\end{deluxetable*}

\subsection{New Photometric LRD Candidates in DJA} \label{subsec:dja-lrd}

We apply the classifier to the \texttt{grizli} v7 NIRCam mosaics from all 36 public fields in the Dawn JWST Archive \citep{Heintz2024DJA, deGraaff2025DJA}. To construct a high-quality sample for classification, we require an F444W signal-to-noise ratio greater than 15, an F444W flux density exceeding $0.05\mu\mathrm{Jy}$, and compact source morphology based on the \textsc{SExtractor} \citep{Bertin1996} semi-major axis measurement ($2 \leq a_\mathrm{image} \leq 5$ pixels). For sources with available EAZY photometric redshifts, we further require $3 < z_\mathrm{phot} < 12$. These criteria are designed to retain compact, high-redshift sources with reliable photometry while excluding low-quality detections and extended low-redshift galaxies. The final sample contains 26,151 sources with valid model inputs.

Applying the trained classifier to this sample yields 357 sources with classification probabilities of $p \geq 0.80$, which we identify as high-confidence photometric LRD candidates, of which 218 have $p>0.90$. 
Table~\ref{tab:dja_candidates} presents the 10 highest-probability candidates, while the full catalog is made available electronically. The candidates span a broad range of photometric redshifts, luminosities, and continuum properties, demonstrating that the classifier is capable of recovering LRD-like systems across a wide region of parameter space.

\begin{figure*}
    \centering
    \includegraphics[width=\linewidth]{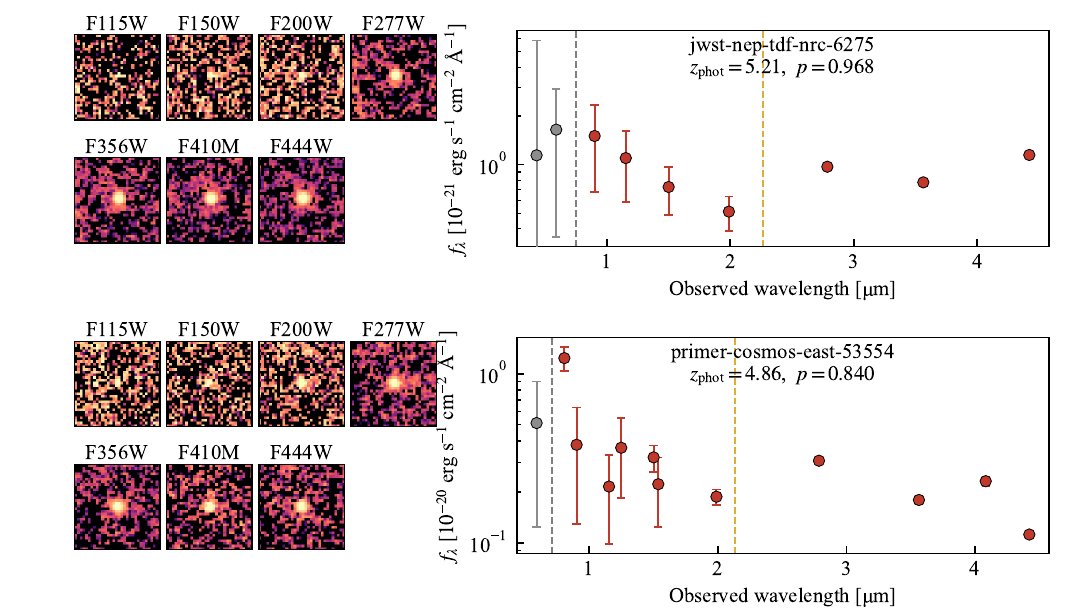}
    \caption{Two candidates from the DJA imaging search. The panels show the seven NIRCam cutouts and broadband SED with uncertainties. Dashed grey and yellow lines represent the expected Ly$\alpha$ and Balmer-break wavelengths, respectively.}
    \label{fig:dja_showcase}
\end{figure*}

Figure~\ref{fig:dja_showcase} presents two representative examples from the candidate sample. For each source, we show both the multi-band NIRCam imaging and the corresponding SED constructed from the available photometry. Both candidates exhibit compact red morphologies and pronounced V-shaped SEDs, characterized by a blue rest-frame ultraviolet continuum and a much redder optical continuum. These features closely resemble those observed in spectroscopically confirmed LRDs and illustrate the ability of the image-based classifier to identify promising LRD candidates using imaging data alone. More broadly, the large number of newly identified candidates demonstrates the potential of representation-based searches to expand current LRD samples beyond those accessible through existing spectroscopic compilations and traditional color-based selection methods.

\begin{deluxetable*}{rlccccccccc}
\tablecaption{Photometric properties of the selected DJA LRD candidates. \label{tab:dja_candidates}}
\tablehead{\colhead{\#} & \colhead{ID} & \colhead{RA} & \colhead{Dec} & \colhead{$z_\mathrm{phot}$} & \colhead{$p$} & \colhead{$\beta_\mathrm{UV}$} & \colhead{$\beta_\mathrm{opt}$} & \colhead{F150W$-$F200W} & \colhead{F277W$-$F444W} & \colhead{$m_\mathrm{F444W}$}}
\startdata
1 & gdn-5343 & 189.26314 & 62.15123 & 5.56 & 0.970 & -2.07 & -0.02 & -0.05 & 1.07 & 26.84 \\
2 & ceers-par-64561 & 214.89998 & 52.77001 & \nodata & 0.970 & \nodata & \nodata & \nodata & \nodata & 25.94 \\
3 & ceers-par-37972 & 214.93977 & 52.73186 & \nodata & 0.968 & \nodata & \nodata & 0.17 & 0.68 & 26.01 \\
4 & jwst-nep-tdf-nrc-6275 & 260.68276 & 65.76421 & 5.21 & 0.968 & -1.38 & -0.95 & 0.23 & 1.18 & 26.72 \\
5 & plck-g165p67-7792 & 171.79576 & 42.48905 & \nodata & 0.968 & \nodata & \nodata & 0.12 & 0.84 & 25.17 \\
6 & abell2744par-11433 & 3.64719 & -30.48121 & \nodata & 0.967 & \nodata & \nodata & 0.28 & 1.04 & 26.33 \\
7 & primer-uds-south-54098 & 34.46688 & -5.23115 & 3.54 & 0.966 & -2.63 & 0.80 & 0.66 & 0.86 & 24.85 \\
8 & primer-cosmos-east-57507 & 150.15047 & 2.37787 & 5.31 & 0.966 & -1.69 & 1.09 & 0.15 & 1.25 & 26.52 \\
9 & elgordo-3136 & 15.76107 & -49.24605 & 4.47 & 0.966 & -1.60 & 1.36 & -0.06 & 0.68 & 25.19 \\
10 & primer-cosmos-east-51262 & 150.20411 & 2.35546 & 7.13 & 0.966 & -0.14 & \nodata & \nodata & 1.95 & 24.76 \\
\enddata
\tablecomments{We use \nodata\ to indicate fields without a photometric-redshift catalog or a slope that cannot be constrained by the available bands. The full table contains 357 sources and is published in its entirety in machine-readable format.}
\end{deluxetable*}

\subsection{Applications to Additional JWST Imaging Surveys} \label{subsec:external-lrd}


We apply the trained classifier to the SAPPHIRES and BEACON imaging datasets to explore its utility for identifying LRD candidates in additional JWST surveys. These applications are intended to demonstrate the transferability of the model and do not provide independent measurements of classification completeness or purity.

The first dataset is the early data release of the SAPPHIRES survey \citep{Sun2025}, which provides deep JWST/NIRCam imaging of the lensing galaxy cluster MACS~J0416.1$-$2403 over a total area of $16.9\ \mathrm{arcmin}^2$. The survey includes observations in 13 NIRCam filters spanning 0.6--5.0~$\mu$m and reaches point-source depths of approximately 28.4-29.5 AB mag. Applying the same classification threshold ($p \geq 0.80$) and subsequent image-quality review yields five LRD candidates. Notably, one of these sources matches the previously reported LRD candidate ID-8760 identified by \citet{Fudamoto2025}, which they propose as a potential low-mass AGN or LRD at $z=8.47$. Our classifier independently recovers this source with $p=0.889$. As summarized in Table~\ref{tab:sapphires}, the five candidates exhibit compact morphologies and red NIRCam colors, with $F277W-F444W$ colors ranging from 0.5 to 2.7 mag, consistent with the broad photometric properties of previously identified LRDs.

\begin{deluxetable*}{rlccccccccc}
\tablecaption{Photometric properties of the $5$ SAPPHIRES LRD candidates selected with $p\geq0.8$. \label{tab:sapphires}}
\tablehead{\colhead{\#} & \colhead{ID} & \colhead{RA} & \colhead{Dec} & \colhead{$z$} & \colhead{$p$} & \colhead{$\beta_\mathrm{UV}$} & \colhead{$\beta_\mathrm{opt}$} & \colhead{F150W$-$F200W} & \colhead{F277W$-$F444W} & \colhead{$m_\mathrm{F444W}$}}
\startdata
1 & 12717 & 64.014209 & -24.158758 & 4.34 & 0.961 & -0.90 & 0.37 & 0.43 & 0.49 & 26.07 \\
2 & 16584 & 64.003087 & -24.150567 & 6.29 & 0.950 & -0.76 & 2.10 & 0.37 & 2.18 & 26.17 \\
3 & 26160 & 64.004005 & -24.113365 & 4.00 & 0.907 & -1.85 & 2.37 & 0.18 & 0.59 & 27.25 \\
4 & 8760 & 63.957333 & -24.167468 & 8.48 & 0.889 & -1.00 & 3.00 & -0.10 & 1.77 & 26.14 \\
5 & 14602 & 63.948942 & -24.154661 & 7.80 & 0.868 & -2.84 & 1.09 & \nodata & 2.67 & 27.24 \\
\enddata
\tablecomments{Again, we use \nodata\ to indicate quantities that cannot be reliably measured because the available photometric bands do not adequately constrain the corresponding color or continuum slope.}
\end{deluxetable*}

The second dataset is BEACON DR1, a JWST Cycle~2 pure-parallel survey comprising 19 independent fields and covering approximately $0.3\ \mathrm{deg}^2$ \citep{Morishita2025}. After applying the same preprocessing and quality-selection procedures, 3327 sources remain with valid model inputs. Of these, 18 are assigned probabilities exceeding $p=0.80$ (Table~\ref{tab:beacon_candidates}). BEACON 0014-3025-6766 is the LRD uncover-61-v4-27114 \citep{Barro2026}.


The applications to SAPPHIRES and BEACON yield plausible LRD candidates under substantially different observing conditions and independently recover one previously reported candidate in each survey. These results demonstrate the practical applicability and transferability of the selection pipeline across diverse JWST imaging datasets. However, the recovered candidates and their numbers should not be interpreted as measures of classification completeness. A meaningful recovery fraction would require a complete census of previously known LRDs in each survey, together with an assessment of which sources satisfy the classifier's input and quality-selection requirements.  

\begin{deluxetable*}{rlccccccccc}
\tabletypesize{\scriptsize}
\tablecaption{Photometric properties of the $18$ BEACON LRD candidates selected at $p\geq0.8$. \label{tab:beacon_candidates}}
\tablehead{\colhead{\#} & \colhead{ID} & \colhead{RA} & \colhead{Dec} & \colhead{$z_\mathrm{phot}$} & \colhead{$p$} & \colhead{$\beta_\mathrm{UV}$} & \colhead{$\beta_\mathrm{opt}$} & \colhead{F115W$-$F200W} & \colhead{F277W$-$F444W} & \colhead{$m_\mathrm{F444W}$}}
\startdata
1 & beacon\_0217-0504-14370 & 34.32355 & -5.09227 & 11.03 & 0.929 & 0.31 & \nodata & \nodata & 1.70 & 26.49 \\
2 & beacon\_0332-2749-22659 & 53.00039 & -27.80568 & 4.45 & 0.897 & -2.09 & 1.69 & 1.30 & 0.86 & 23.82 \\
3 & beacon\_0959p0200-7853 & 149.84087 & 2.01361 & 6.97 & 0.888 & -0.82 & 1.82 & 0.69 & 2.47 & 25.43 \\
4 & beacon\_0959p0200-14387 & 149.84462 & 2.03539 & 4.63 & 0.883 & -2.93 & 1.40 & 1.30 & 0.02 & 21.51 \\
5 & beacon\_0502-4338-1276 & 75.45445 & -43.64950 & 4.52 & 0.869 & 0.57 & 0.70 & 1.52 & 0.58 & 24.37 \\
6 & beacon\_2316-5910-27047 & 349.08880 & -59.14788 & 5.42 & 0.868 & -1.74 & 1.39 & 0.15 & 1.73 & 25.68 \\
7 & beacon\_0447-2637-5821 & 71.68219 & -26.61793 & 7.17 & 0.865 & -1.21 & 1.91 & -0.90 & 1.70 & 25.70 \\
8 & beacon\_0332-2745-15718 & 53.04577 & -27.73467 & 3.03 & 0.865 & -1.04 & \nodata & \nodata & 0.92 & 26.31 \\
9 & beacon\_0217-0504-4690 & 34.30628 & -5.09088 & 7.14 & 0.860 & -0.58 & 1.94 & \nodata & 2.61 & 26.72 \\
10 & beacon\_1420p5252-7960 & 215.09023 & 52.87087 & 3.96 & 0.857 & -1.91 & -0.82 & 0.66 & 1.72 & 26.66 \\
11 & beacon\_0332-2749-7936 & 53.00050 & -27.82327 & 3.69 & 0.853 & -1.79 & 0.10 & 0.85 & 0.62 & 25.60 \\
12 & beacon\_0502-4338-17372 & 75.44405 & -43.63003 & 6.34 & 0.844 & -0.80 & 1.57 & 0.71 & 1.82 & 24.91 \\
13 & beacon\_0014-3025-6766 & 3.54584 & -30.39570 & 5.11 & 0.831 & -1.47 & 1.31 & 0.45 & 1.68 & 25.34 \\
14 & beacon\_0217-0508-14889 & 34.28143 & -5.10287 & 5.10 & 0.816 & -0.12 & 1.41 & \nodata & 1.80 & 26.04 \\
15 & beacon\_2316-5910-13989 & 349.16520 & -59.19123 & 7.20 & 0.814 & -0.93 & 0.51 & 0.42 & 2.15 & 23.45 \\
16 & beacon\_0442-4613-7155 & 70.45215 & -46.22994 & 4.06 & 0.814 & -9.05 & \nodata & \nodata & -0.12 & 21.80 \\
17 & beacon\_0217-0509-4499 & 34.23260 & -5.15447 & 3.23 & 0.808 & \nodata & \nodata & \nodata & -0.28 & 21.64 \\
18 & beacon\_2316-5910-16796 & 349.09050 & -59.16090 & 5.50 & 0.805 & -1.48 & 1.35 & 0.31 & 1.71 & 25.03 \\
\enddata
\tablecomments{We use \nodata\ to indicate cases where the available photometric bands do not provide sufficient wavelength coverage, or where one or more required flux measurements are missing, non-finite, or non-positive, preventing a reliable determination of the corresponding color or continuum slope.}
\end{deluxetable*}

\section{Discussions}
\label{sec:discussion}

Recovering previously reported candidates, along with identifying new sources, shows that the classifier can be effectively applied to additional JWST datasets beyond those used for training. The selected candidates exhibit photometric and morphological
properties broadly consistent with those of known LRDs. This provides
a useful consistency check, although survey-dependent selection effects
may also contribute to the observed similarities. At the same time, variations in survey depth, filter coverage, source extraction, and data reduction procedures can affect the quality and interpretation of the input images. Consequently, survey-specific input validation and visual inspection remain important components of the candidate selection process. Owing to data limitations, not all the newly identified LRD candidates possess valid photometric redshifts or complete photometric measurements. We therefore restrict our analysis to those sources for which all relevant measurements are available below.

\begin{figure}
    \centering
    \includegraphics[width=\columnwidth]{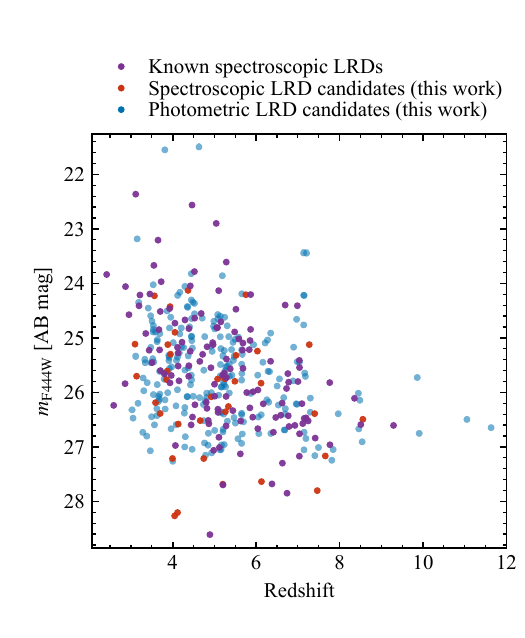}
    \caption{F444W AB magnitude versus redshift for known spectroscopic LRDs and model-selected candidates. Spectroscopic redshifts are used where available and photometric redshifts otherwise.}
    \label{fig:z-f444w-candidates}
\end{figure}

Figure~\ref{fig:z-f444w-candidates} compares the redshift and observed F444W magnitude distributions of spectroscopically confirmed LRDs and the candidates identified through the spectroscopic, DJA, SAPPHIRES, and BEACON searches. The newly identified candidates occupy broadly the same region of the $z$--$m_{\rm F444W}$ parameter space as the confirmed LRD population, indicating that the classifier is recovering sources with observational properties consistent with known LRDs.

\subsection{Selection boundaries and inferred LRD population} \label{subsec:selection-boundaries}

\begin{figure}
    \centering
    \includegraphics[width=\linewidth]{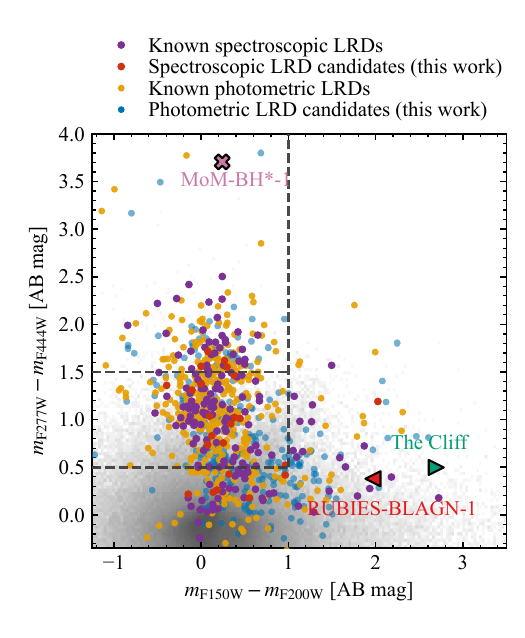}
    \caption{Observed colors of the LRD samples. The grey density shows the DJA photometric population, and colored symbols denote the samples listed in the legend. The Cliff \citep{deGraaff2025b}, MoM-BH*-1 \citep{Naidu2025}, and RUBIES-BLAGN-1 \citep{Wangbingjie2025} are marked separately. Dashed lines mark the commonly adopted LRD color-selection boundaries at $m_{\mathrm{F150W}}-m_{\mathrm{F200W}}=1.0$ and $m_{\mathrm{F277W}}-m_{\mathrm{F444W}}=0.5$ and $1.5$ mag.}
    \label{fig:selection_space}
\end{figure}

The inferred abundance and observed-property distributions of LRDs depend on the selection criteria used to define the sample \citep[e.g.,][]{Rinaldi2026}. Figure~\ref{fig:selection_space} compares the spectroscopic and photometric LRD candidates identified by our classifier with literature LRD samples in the observed $F150W-F200W$ versus $F277W-F444W$ color plane. The literature samples and our candidates occupy broadly overlapping regions of color space, indicating that the classifier successfully recovers sources with photometric properties similar to previously identified LRDs. At the same time, the model-selected candidates exhibit a substantially wider distribution, extending beyond the regions traditionally associated with LRD selection, particularly toward bluer $F277W-F444W$ colors. This behavior highlights a key strength of the image-based classifier: by leveraging the full multi-band imaging information rather than predefined color boundaries, it can probe candidate populations outside fixed color-selection boundaries.

The existence of spectroscopically confirmed or well-studied LRDs outside the conventional color-selection regions further supports this interpretation. Examples such as The Cliff \citep{deGraaff2025}, MoM-BH*-1 \citep{Naidu2025}, and RUBIES-BLAGN-1 \citep{Wangbingjie2025} occupy distinct locations in the color plane yet are widely recognized as members of the LRD population. These objects exhibit characteristic LRD features, including strong Balmer breaks and red continua, demonstrating that LRDs do not occupy a single, well-defined color locus. Instead, the observed color distribution likely reflects a combination of intrinsic diversity in stellar populations, dust attenuation, emission-line strengths, AGN contributions, and redshift evolution.

The recovery of candidates in these extended regions suggests that current color-selected samples may not fully capture the diversity of the LRD population. This possibility is also consistent with recent studies that have adopted broader or alternative color-selection criteria and identified LRD candidates spanning a wider range of photometric properties \citep{Rinaldi2026}. Nevertheless, photometric properties alone are insufficient to establish the nature of these sources. Spectroscopic follow-up will be required to determine whether they represent genuine LRDs or contaminants from other high-redshift galaxy populations and Galactic stars. If a substantial fraction are confirmed, the results would indicate that LRDs span a broader range of observed colors than currently represented in existing samples, highlighting the potential of image-based selection to uncover previously overlooked members of the population.

\subsection{Photometric diagnostics of ML-selected candidates} \label{subsec:photometric-diagnostics}

\begin{figure}
    \centering
    \includegraphics[width=\linewidth]{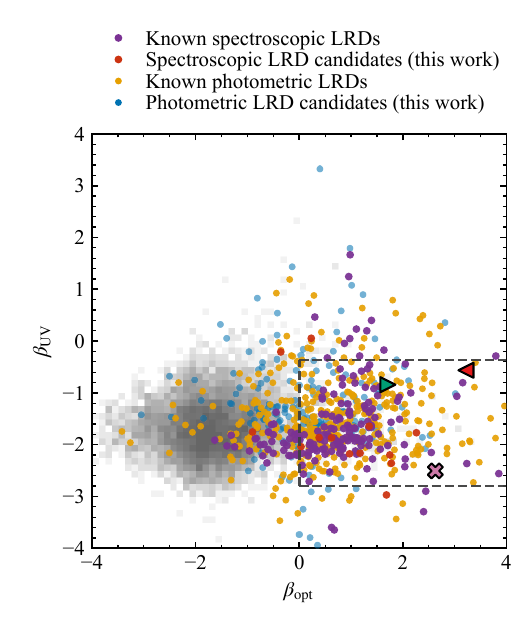}
    \caption{Rest-frame optical and UV continuum slopes. The grey density shows the DJA photometric population, and colored symbols denote the samples listed in the legend. The Cliff \citep{deGraaff2025b}, MoM-BH*-1 \citep{Naidu2025}, and RUBIES-BLAGN-1 \citep{Wangbingjie2025} are marked separately. Dashed lines mark the conventional V-shaped continuum region defined by $\beta_{\rm opt}>0$ and $-2.8<\beta_{\rm UV}<-0.37$.}
    \label{fig:beta_beta}
\end{figure}

Data limitations restrict reliable $\beta_{\rm UV}$ and $\beta_{\rm opt}$ estimates to 189 of the 357 candidates, which meet our adopted slope-quality criteria and are presented here. The $\beta_{\rm opt}$--$\beta_{\rm UV}$ plane provides a complementary view of the characteristic V-shaped continuum that defines the LRD population. Following \citet{deGraaff2025}, we measure continuum slopes from NIRSpec/PRISM spectra for spectroscopic sources and from photometric bands sampling the rest-frame 1216--3646~\AA\ and 3500--7000~\AA\ wavelength ranges for imaging-selected sources. Figure~\ref{fig:beta_beta} compares the resulting distributions for previously known LRDs and the candidates identified in this work. The newly identified candidates broadly overlap the locus occupied by known LRDs, indicating that they share the characteristic combination of blue UV continua and substantially redder optical continua associated with strong Balmer-break systems. At the same time, both the spectroscopic and photometric candidates extend beyond the conventional concentration of sources in this plane, spanning a wider range of $\beta_{\rm UV}$ and $\beta_{\rm opt}$ values. 

The extended distribution may reflect a combination of intrinsic diversity within the LRD population and measurement uncertainties. In particular, photometrically derived continuum slopes are sensitive to filter coverage, photometric-redshift uncertainties, and contamination from strong emission lines, all of which can broaden the observed distribution. Consequently, sources lying outside the region traditionally associated with LRDs should not necessarily be interpreted as a distinct population. Rather, they represent promising candidates that can help test whether the observed diversity of LRD continua is larger than currently inferred from spectroscopically confirmed samples.

While the overlap with known LRDs supports the validity of the candidate sample, a clearer separation between LRDs and the general high-redshift galaxy population will require both larger spectroscopic training sets and targeted follow-up observations. Such data will be particularly important for determining whether sources with extreme Balmer-break signatures represent previously under-sampled regions of the LRD population or arise from contamination within current photometric selections.

\subsection{Limitations}

Despite its promising performance, several limitations should be considered when interpreting the results of this work. First, the training sample remains relatively small and potentially incomplete. The positive class is constructed from currently available spectroscopically confirmed LRDs, which number only a few hundred objects and may not fully capture the diversity of the underlying population. Similarly, the numbers of spectroscopically confirmed brown dwarfs and high-redshift quiescent galaxies available for training are also limited, restricting the model's ability to learn the full range of properties exhibited by these potentially important contaminants. Moreover, as demonstrated by the spectroscopic candidates identified in this work, some genuine LRDs are likely still labeled as ordinary high-redshift galaxies in existing catalogs. Such label incompleteness introduces uncertainty into both the training process and the evaluation metrics, potentially biasing the learned decision boundary and causing the reported precision to underestimate the true performance of the classifier. As larger and more representative spectroscopic samples become available, retraining with improved labels will not only enhance model robustness and reduce selection biases, but will also enable a more complete exploration of the diversity of the LRD population and its principal contaminants.

A second consideration concerns the input data used by the classifier. Our model relies exclusively on JWST/NIRCam imaging in seven broad-band filters, which are not available for every source or survey. Consequently, it does not exploit complementary information from spectroscopy, medium-band photometry, variability, or longer-wavelength observations. Moreover, variations in filter availability alter the rest-frame wavelength coverage and may affect the visibility of features such as the UV continuum slope and Balmer break. Incorporating more diverse training data and complementary multi-wavelength information, together with architectures that better accommodate missing filters, may further improve the robustness and applicability of the method. Missing filters are represented by zero-valued image channels, which could potentially encode survey-specific observing patterns. We therefore examined classifier performance as a function of filter availability across the labeled sample. Labeled LRDs are recovered across the five-, six-, and seven-band subsets, with recall broadly consistent with that of the full sample, suggesting that the classifier can operate effectively under heterogeneous filter coverage. However, the limited number of labeled LRDs in the incomplete-filter subsets and the likely incompleteness of existing catalogs prevent a definitive assessment of filter-dependent effects. In addition, although applications to independent surveys demonstrate the feasibility of transferring the classifier across datasets, differences in filter coverage, survey depth, image quality, source extraction, and photometric processing mean that cross-survey completeness and purity remain to be established. Larger, more uniformly characterized spectroscopic samples will be important for further testing the robustness of the method.

Finally, the current classifier is designed as a candidate-selection tool, and its performance is naturally linked to the characteristics of the training data and input representations. The model is optimized to identify sources that resemble known LRDs in the learned image-embedding space and therefore may not recover all manifestations of the population. Objects with properties that are underrepresented in the training sample may receive lower classification probabilities even if they are physically related to LRDs. In addition, the learned representations are not directly interpretable in terms of physical quantities such as stellar age, dust attenuation, AGN contribution, or Balmer-break strength. Future work combining imaging, spectroscopy, and physically motivated latent representations may help improve both classification performance and physical interpretability.

\section{Conclusion} \label{sec:conclusion}

In this work, we develop a spectroscopy-informed deep-learning framework for identifying Little Red Dots from JWST imaging data, trained on 127 spectroscopically confirmed LRDs and $\sim7,500$ literature-labeled non-LRD sources. By aligning NIRCam images and NIRSpec/PRISM spectra during training, the model learns spectroscopically informed image representations while requiring only NIRCam imaging at inference. On a held-out validation sample, it achieves a recall of 0.92 and a precision of 0.79 at $p=0.80$, demonstrating that imaging data alone can recover the majority of currently known LRDs.

The classifier identifies 35 LRD candidates among spectroscopically observed high-redshift galaxies that were previously labeled as non-LRDs. Many of these sources exhibit independent characteristics commonly associated with LRDs. Among the 13 candidates with usable medium- or high-resolution H$\alpha$ spectroscopy, 10 show a broad H$\alpha$ component with FWHM values of approximately $1100-3500\ \mathrm{km\ s^{-1}}$. In addition, most candidates display compact red NIRCam morphologies and pronounced V-shaped PRISM continua. These results provide independent support for the LRD interpretation of many model-selected candidates and suggest that a significant fraction of the apparent false positives may correspond to previously uncatalogued members of the LRD population.

We further apply the classifier to public JWST imaging surveys. The model identifies 357 previously uncatalogued candidates across 32 fields in the Dawn JWST Archive, together with five candidates in SAPPHIRES and eighteen in BEACON. 

The resulting candidates occupy broadly the same regions of redshift, magnitude, color, and continuum-slope parameter space as previously known LRDs, while extending beyond several commonly adopted color-selection boundaries. This suggests that the observed diversity of the LRD population may not be fully captured by the current color-based selections and highlights the potential of image-based classification to identify previously overlooked candidates. Nevertheless, continuum-slope analysis and spectroscopic inspection indicate that contamination from compact quiescent galaxies and other strong-Balmer-break systems remains present, underscoring the importance of spectroscopic follow-up for establishing the nature of individual candidates and quantifying the completeness and purity of future LRD samples.

Several factors currently limit a direct interpretation of the candidate counts as measurements of the intrinsic LRD population. These include the small number of spectroscopically confirmed LRDs available for training and validation, incompleteness in existing catalog labels, heterogeneous imaging inputs across JWST surveys, and uncertainties in probability calibration. While the candidate samples substantially expand the current inventory of potential LRDs, they should not be interpreted as a complete census of the population. Future progress will benefit from larger spectroscopic training sets, improved catalog completeness, and continued follow-up observations. Such efforts will enable more robust characterization of image-selected LRD samples and provide new insights into the diversity and physical nature of this emerging population.

More broadly, this work demonstrates the potential of spectroscopy-informed image classification as a powerful new tool for studying LRDs in the JWST era and future wide-area imaging surveys. By leveraging spectroscopic information during training while requiring only imaging data at inference, the framework enables efficient searches over survey volumes far larger than those currently accessible through spectroscopy alone and reduces reliance on predefined color-selection criteria. The resulting candidate samples suggest that the diversity of the LRD population may extend beyond that represented in existing catalogs. As spectroscopic samples continue to grow and future surveys expand the available imaging footprint, increasingly complete and uniformly selected LRD catalogs will provide new opportunities to investigate the demographics, physical origins, and evolutionary role of LRDs in the early Universe.

\section{Acknowledgments}

C.Z., H.Z., and Q.W. acknowledge support for this work from the National Natural Science Foundation of China (Grant Nos. 12303007, 12533005, and 12233007), the China Manned Space Program (Grant Nos. CMS-CSST-2025-A06 and CMS-CSST-2025-A07) and the financial support from the start-up funding of Huazhong University of Science and Technology.

The data products presented herein were retrieved from the Dawn JWST Archive (DJA). DJA is an initiative of the Cosmic Dawn Center (DAWN), which is funded by the Danish National Research Foundation under grant DNRF140. We acknowledge the many developers and contributors to the JWST Science Calibration Pipeline.


\appendix

\section{Spectra of the High-Probability Spectroscopic LRD Candidates}
\label{app:fp-spectra}
\setcounter{figure}{0}
\renewcommand{\thefigure}{\thesection\arabic{figure}}
\renewcommand{\theHfigure}{appendix.\thesection.\arabic{figure}}

\begin{figure*}[p]
\centering
\includegraphics[width=\textwidth,height=0.88\textheight,keepaspectratio]{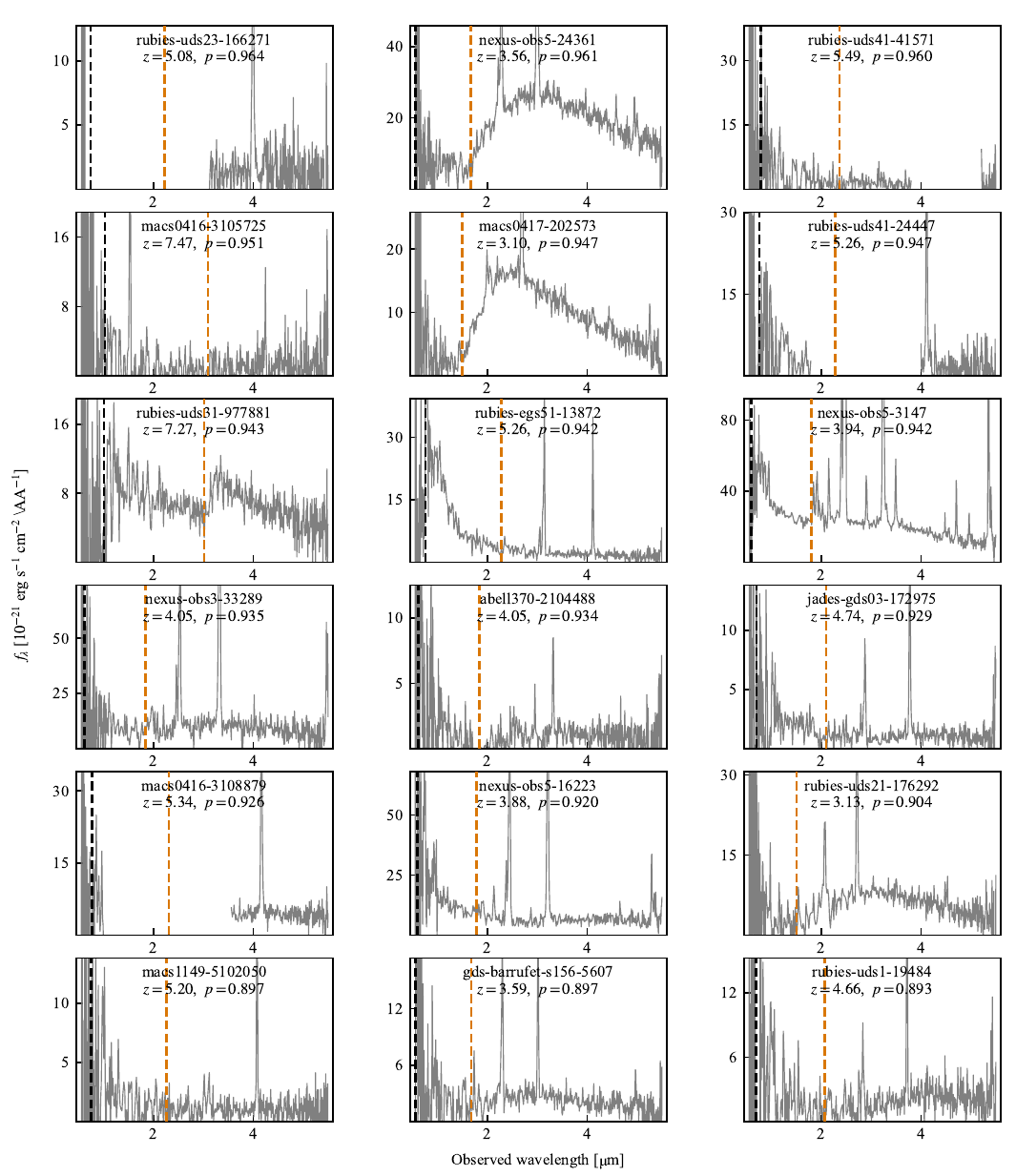}
\caption{Observed-frame NIRSpec/PRISM spectra of candidates in Table~\ref{tab:fp_recovery}. Dashed lines mark the expected Ly$\alpha$ wavelength and Balmer break.}
\label{fig:fp_spectra_appendix_01}
\end{figure*}

\begin{figure*}[p]
\centering
\includegraphics[width=\textwidth,height=0.88\textheight,keepaspectratio]{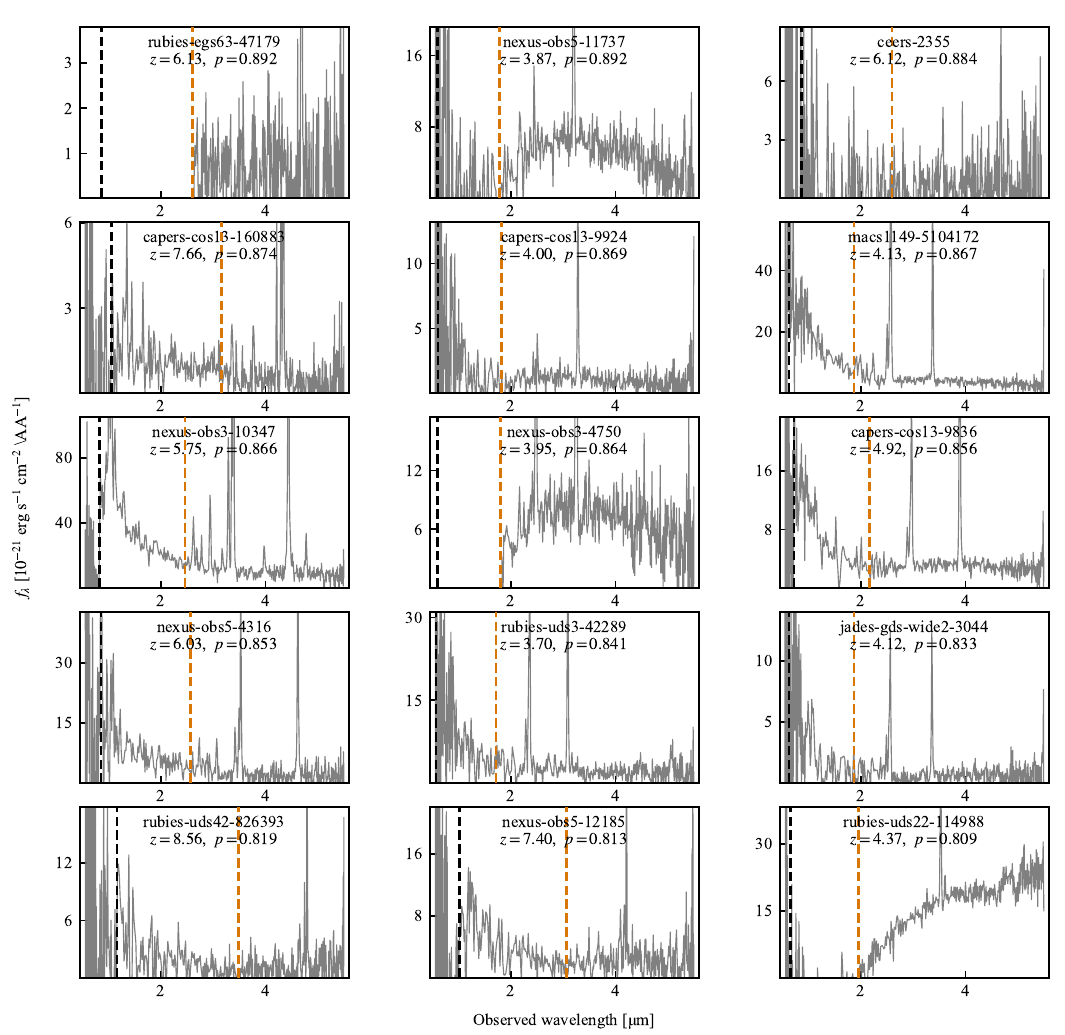}
\caption{Same as figure~\ref{fig:fp_spectra_appendix_01}, showing the rest of candidates in Table~\ref{tab:fp_recovery}.}
\label{fig:fp_spectra_appendix_02}
\end{figure*}

Figures~\ref{fig:fp_spectra_appendix_01} and \ref{fig:fp_spectra_appendix_02} present the observed-frame NIRSpec/PRISM spectra of the remaining 33 candidates discussed in Section~\ref{subsec:spec-lrd}, ordered by decreasing model probability.

\clearpage

\bibliography{ref.bib}{}

@ARTICLE{Ye2024,
       author = {{Ye}, Guangping and {Zhang}, Huanian and {Wu}, Qingwen},
        title = "{Machine Learning{\textendash}based Search of High-redshift Quasars}",
      journal = {\apjs},
         year = 2024,
        month = nov,
       volume = {275},
       number = {1},
          eid = {19},
        pages = {19},
          doi = {10.3847/1538-4365/ad79ee},
archivePrefix = {arXiv},
       eprint = {2409.02167},
 primaryClass = {astro-ph.GA},
       adsurl = {https://ui.adsabs.harvard.edu/abs/2024ApJS..275...19Y}
}

@ARTICLE{ZhangC2026b,
       author = {{Zhang}, Chenxuan and {Zhang}, Huanian and {Wu}, Qingwen and {Ho}, Luis C. and {Wang}, Jian-Min},
        title = "{Reduced incidence of little red dots at $z < 3$ based on number density and halo mass evolution}",
      journal = {Nature Astronomy},
         year = 2026,
        month = oct,
          doi = {10.1038/s41550-026-03001-6},
          url = {https://www.nature.com/articles/s41550-026-03001-6}
}

@ARTICLE{Matthee2026,
       author = {{Matthee}, Jorryt and {Torralba}, Alberto and {Pezzulli}, Gabriele and {Naidu}, Rohan P. and {Chisholm}, John and {Mascia}, Sara and {Greene}, Jenny E. and {Ishikawa}, Yuzo and {Gronke}, Max and {Wuyts}, Stijn and {Bordoloi}, Rongmon and {Brammer}, Gabriel and {Chang}, Seok-Jun and {Eilers}, Anna-Christina and {de Graaff}, Anna and {Hviding}, Raphael E. and {Iani}, Edoardo and {Illingworth}, Garth and {Kashino}, Daichi and {Labbe}, Ivo and {Ma}, Yilun and {Maseda}, Michael V. and {Meyer}, Romain and {Nelson}, Erica and {Oesch}, Pascal and {Xiao}, Mengyuan},
        title = "{The Engine and its Flows: Little Red Dot spectra are shaped by the column densities of their gas envelopes}",
      journal = {arXiv e-prints},
         year = 2026,
        month = mar,
          eid = {arXiv:2603.17667},
        pages = {arXiv:2603.17667},
          doi = {10.48550/arXiv.2603.17667},
archivePrefix = {arXiv},
       eprint = {2603.17667},
 primaryClass = {astro-ph.GA},
       adsurl = {https://ui.adsabs.harvard.edu/abs/2026arXiv260317667M}
}

@ARTICLE{ZhangHN2025,
       author = {{Zhang}, Huanian and {Ye}, Guangping and {Wu}, Rongyu and {Zaritsky}, Dennis},
        title = "{A Local Dwarf Galaxy Search Using Machine Learning}",
      journal = {\apjs},
         year = 2025,
        month = may,
       volume = {278},
       number = {1},
          eid = {18},
        pages = {18},
          doi = {10.3847/1538-4365/adbf0a},
archivePrefix = {arXiv},
       eprint = {2503.00109},
 primaryClass = {astro-ph.GA},
       adsurl = {https://ui.adsabs.harvard.edu/abs/2025ApJS..278...18Z}
}

@ARTICLE{deGraaff2025,
       author = {{de Graaff}, Anna and {Hviding}, Raphael E. and {Naidu}, Rohan P. and {Greene}, Jenny E. and {Miller}, Tim B. and {Leja}, Joel and {Matthee}, Jorryt and {Brammer}, Gabriel and {Katz}, Harley and {Bezanson}, Rachel and {Boogaard}, Leindert A. and {Bose}, Sownak and {Chisholm}, John and {Cleri}, Nikko J. and {Dayal}, Pratika and {Feldmann}, Robert and {Fudamoto}, Yoshinobu and {Fujimoto}, Seiji and {Furtak}, Lukas J. and {Glazebrook}, Karl and {Gottumukkala}, Rashmi and {Heintz}, Kasper E. and {Kokorev}, Vasily and {Labbe}, Ivo and {Maseda}, Michael V. and {McConachie}, Ian and {Nanayakkara}, Themiya and {Nelson}, Erica and {Nowaczyk}, Przemys{\l}aw and {Oesch}, Pascal A. and {Rix}, Hans-Walter and {Setton}, David J. and {Torralba}, Alberto and {Walter}, Fabian and {Wang}, Bingjie and {Weibel}, Andrea and {van der Wel}, Arjen},
        title = "{Little Red Dots host Black Hole Stars: A unified family of gas-reddened AGN revealed by JWST/NIRSpec spectroscopy}",
      journal = {arXiv e-prints},
         year = 2025,
        month = nov,
          eid = {arXiv:2511.21820},
        pages = {arXiv:2511.21820},
          doi = {10.48550/arXiv.2511.21820},
archivePrefix = {arXiv},
       eprint = {2511.21820},
 primaryClass = {astro-ph.GA},
       adsurl = {https://ui.adsabs.harvard.edu/abs/2025arXiv251121820D}
}

@ARTICLE{Hviding2025,
       author = {{Hviding}, Raphael E. and {de Graaff}, Anna and {Miller}, Tim B. and {Setton}, David J. and {Greene}, Jenny E. and {Labb{\'e}}, Ivo and {Brammer}, Gabriel and {Bezanson}, Rachel and {Boogaard}, Leindert A. and {Cleri}, Nikko J. and {Leja}, Joel and {Maseda}, Michael V. and {McConachie}, Ian and {Matthee}, Jorryt and {Naidu}, Rohan P. and {Oesch}, Pascal A. and {Wang}, Bingjie and {Whitaker}, Katherine E. and {Williams}, Christina C.},
        title = "{RUBIES: A spectroscopic census of little red dots: All point sources with v-shaped continua have broad lines}",
      journal = {\aap},
         year = 2025,
        month = oct,
       volume = {702},
          eid = {A57},
        pages = {A57},
          doi = {10.1051/0004-6361/202555816},
archivePrefix = {arXiv},
       eprint = {2506.05459},
 primaryClass = {astro-ph.GA},
       adsurl = {https://ui.adsabs.harvard.edu/abs/2025A&A...702A..57H}
}

@ARTICLE{Zhangchenxuan2025,
       author = {{Zhang}, Chenxuan and {Wu}, Qingwen and {Fan}, Xiao and {Ho}, Luis C. and {Wu}, Jiancheng and {Zhang}, Huanian and {Lyu}, Bing and {Cao}, Xinwu and {Wang}, Jianmin},
        title = "{The composite spectrum of little red dots from a standard inner disk and an unstable outer disk}",
      journal = {Nature Astronomy},
         year = 2026,
        month = may,
       volume = {10},
        pages = {753-761},
          doi = {10.1038/s41550-026-02785-x},
archivePrefix = {arXiv},
       eprint = {2505.12719},
 primaryClass = {astro-ph.HE},
       adsurl = {https://ui.adsabs.harvard.edu/abs/2026NatAs..10..753Z}
}

@ARTICLE{Kocevski2025,
       author = {{Kocevski}, Dale D. and {Finkelstein}, Steven L. and {Barro}, Guillermo and {Taylor}, Anthony J. and {Calabr{\`o}}, Antonello and {Laloux}, Brivael and {Buchner}, Johannes and {Trump}, Jonathan R. and {Leung}, Gene C.~K. and {Yang}, Guang and {Dickinson}, Mark and {P{\'e}rez-Gonz{\'a}lez}, Pablo G. and {Pacucci}, Fabio and {Inayoshi}, Kohei and {Somerville}, Rachel S. and {McGrath}, Elizabeth J. and {Akins}, Hollis B. and {Bagley}, Micaela B. and {Bowler}, Rebecca A.~A. and {Bisigello}, Laura and {Carnall}, Adam and {Casey}, Caitlin M. and {Cheng}, Yingjie and {Cleri}, Nikko J. and {Costantin}, Luca and {Cullen}, Fergus and {Davis}, Kelcey and {Donnan}, Callum T. and {Dunlop}, James S. and {Ellis}, Richard S. and {Ferguson}, Henry C. and {Fujimoto}, Seiji and {Fontana}, Adriano and {Giavalisco}, Mauro and {Grazian}, Andrea and {Grogin}, Norman A. and {Hathi}, Nimish P. and {Hirschmann}, Michaela and {Huertas-Company}, Marc and {Holwerda}, Benne W. and {Illingworth}, Garth and {Juneau}, St{\'e}phanie and {Kartaltepe}, Jeyhan S. and {Koekemoer}, Anton M. and {Li}, Wenxiu and {Lucas}, Ray A. and {Magee}, Dan and {Mason}, Charlotte and {McLeod}, Derek J. and {McLure}, Ross J. and {Napolitano}, Lorenzo and {Papovich}, Casey and {Pirzkal}, Nor and {Rodighiero}, Giulia and {Santini}, Paola and {Wilkins}, Stephen M. and {Yung}, L.~Y. Aaron},
        title = "{The Rise of Faint, Red Active Galactic Nuclei at z > 4: A Sample of Little Red Dots in the JWST Extragalactic Legacy Fields}",
      journal = {\apj},
         year = 2025,
        month = jun,
       volume = {986},
       number = {2},
          eid = {126},
        pages = {126},
          doi = {10.3847/1538-4357/adbc7d},
archivePrefix = {arXiv},
       eprint = {2404.03576},
 primaryClass = {astro-ph.GA},
       adsurl = {https://ui.adsabs.harvard.edu/abs/2025ApJ...986..126K}
}

@article{Davis2026,
  title={OCEANS of Absorption: High-resolution NIRSpec Spectroscopy Reveals Diverse Balmer-line Absorption in Little Red Dots},
  author={Davis, Kelcey and Brooks, Madisyn and Simons, Raymond C and Trump, Jonathan R and Barro, Guillermo and Haro, Pablo Arrabal and Backhaus, Bren E and Cleri, Nikko J and de la Vega, Alexander and Finkelstein, Steven L and others},
  journal={arXiv preprint arXiv:2606.00258},
  year={2026}
}

@ARTICLE{Kokorev2024,
       author = {{Kokorev}, Vasily and {Caputi}, Karina I. and {Greene}, Jenny E. and {Dayal}, Pratika and {Trebitsch}, Maxime and {Cutler}, Sam E. and {Fujimoto}, Seiji and {Labb{\'e}}, Ivo and {Miller}, Tim B. and {Iani}, Edoardo and {Navarro-Carrera}, Rafael and {Rinaldi}, Pierluigi},
        title = "{A Census of Photometrically Selected Little Red Dots at 4 < z < 9 in JWST Blank Fields}",
      journal = {\apj},
         year = 2024,
        month = jun,
       volume = {968},
       number = {1},
          eid = {38},
        pages = {38},
          doi = {10.3847/1538-4357/ad4265},
archivePrefix = {arXiv},
       eprint = {2401.09981},
 primaryClass = {astro-ph.GA},
       adsurl = {https://ui.adsabs.harvard.edu/abs/2024ApJ...968...38K}
}

@ARTICLE{Zhangzijian2026,
       author = {{Zhang}, Zijian and {Jiang}, Linhua and {Liu}, Weiyang and {Ho}, Luis C. and {Inayoshi}, Kohei},
        title = "{JWST Insights into Narrow-line Little Red Dots}",
      journal = {\apj},
         year = 2026,
        month = feb,
       volume = {998},
       number = {1},
          eid = {170},
        pages = {170},
          doi = {10.3847/1538-4357/ae33c4},
archivePrefix = {arXiv},
       eprint = {2506.04350},
 primaryClass = {astro-ph.GA},
       adsurl = {https://ui.adsabs.harvard.edu/abs/2026ApJ...998..170Z}
}

@ARTICLE{Langeroodi2023,
       author = {{Langeroodi}, Danial and {Hjorth}, Jens},
        title = "{Little Red Dots or Brown Dwarfs? NIRSpec Discovery of Three Distant Brown Dwarfs Masquerading as NIRCam-selected Highly Reddened Active Galactic Nuclei}",
      journal = {\apjl},
         year = 2023,
        month = nov,
       volume = {957},
       number = {2},
          eid = {L27},
        pages = {L27},
          doi = {10.3847/2041-8213/acfeec},
archivePrefix = {arXiv},
       eprint = {2308.10900},
 primaryClass = {astro-ph.GA},
       adsurl = {https://ui.adsabs.harvard.edu/abs/2023ApJ...957L..27L}
}

@ARTICLE{Tuzhijun2025a,
       author = {{Tu}, Zhijun and {Wang}, Shu and {Chen}, Xiaodian and {Liu}, Jifeng},
        title = "{Three Brown Dwarfs Masquerading as High-redshift Galaxies in JWST Observations}",
      journal = {\apj},
         year = 2025,
        month = feb,
       volume = {980},
       number = {2},
          eid = {230},
        pages = {230},
          doi = {10.3847/1538-4357/adaf9f},
archivePrefix = {arXiv},
       eprint = {2501.16648},
 primaryClass = {astro-ph.SR},
       adsurl = {https://ui.adsabs.harvard.edu/abs/2025ApJ...980..230T}
}

@ARTICLE{Tuzhijun2025b,
       author = {{Tu}, Zhijun and {Wang}, Shu and {Chen}, Xiaodian and {Liu}, Jifeng},
        title = "{A Large Sample of JWST/NIRSpec Brown Dwarfs: New Distant Discoveries}",
      journal = {\apjs},
         year = 2025,
        month = dec,
       volume = {281},
       number = {2},
          eid = {49},
        pages = {49},
          doi = {10.3847/1538-4365/ae112e},
archivePrefix = {arXiv},
       eprint = {2510.02026},
 primaryClass = {astro-ph.SR},
       adsurl = {https://ui.adsabs.harvard.edu/abs/2025ApJS..281...49T}
}

@ARTICLE{Allard2012,
       author = {{Allard}, F. and {Homeier}, D. and {Freytag}, B.},
        title = "{Models of very-low-mass stars, brown dwarfs and exoplanets}",
      journal = {Philosophical Transactions of the Royal Society of London Series A},
         year = 2012,
        month = jun,
       volume = {370},
       number = {1968},
        pages = {2765-2777},
          doi = {10.1098/rsta.2011.0269},
archivePrefix = {arXiv},
       eprint = {1112.3591},
 primaryClass = {astro-ph.SR},
       adsurl = {https://ui.adsabs.harvard.edu/abs/2012RSPTA.370.2765A}
}

@ARTICLE{Heintz2024,
       author = {{Heintz}, Kasper E. and {Watson}, Darach and {Brammer}, Gabriel and {Vejlgaard}, Simone and {Hutter}, Anne and {Strait}, Victoria B. and {Matthee}, Jorryt and {Oesch}, Pascal A. and {Jakobsson}, P{\'a}ll and {Tanvir}, Nial R. and {Laursen}, Peter and {Naidu}, Rohan P. and {Mason}, Charlotte A. and {Killi}, Meghana and {Jung}, Intae and {Hsiao}, Tiger Yu-Yang and {Abdurro'uf} and {Coe}, Dan and {Arrabal Haro}, Pablo and {Finkelstein}, Steven L. and {Toft}, Sune},
        title = "{Strong damped Lyman-{\ensuremath{\alpha}} absorption in young star-forming galaxies at redshifts 9 to 11}",
      journal = {Science},
         year = 2024,
        month = may,
       volume = {384},
       number = {6698},
        pages = {890-894},
          doi = {10.1126/science.adj0343},
archivePrefix = {arXiv},
       eprint = {2306.00647},
 primaryClass = {astro-ph.GA},
       adsurl = {https://ui.adsabs.harvard.edu/abs/2024Sci...384..890H}
}

@ARTICLE{deGraaff2025DJA,
       author = {{de Graaff}, Anna and {Brammer}, Gabriel and {Weibel}, Andrea and {Lewis}, Zach and {Maseda}, Michael V. and {Oesch}, Pascal A. and {Bezanson}, Rachel and {Boogaard}, Leindert A. and {Cleri}, Nikko J. and {Cooper}, Olivia R. and {Gottumukkala}, Rashmi and {Greene}, Jenny E. and {Hirschmann}, Michaela and {Hviding}, Raphael E. and {Katz}, Harley and {Labb{\'e}}, Ivo and {Leja}, Joel and {Matthee}, Jorryt and {McConachie}, Ian and {Miller}, Tim B. and {Naidu}, Rohan P. and {Price}, Sedona H. and {Rix}, Hans-Walter and {Setton}, David J. and {Suess}, Katherine A. and {Wang}, Bingjie and {Whitaker}, Katherine E. and {Williams}, Christina C.},
        title = "{RUBIES: A complete census of the bright and red distant Universe with JWST/NIRSpec}",
      journal = {\aap},
         year = 2025,
        month = may,
       volume = {697},
          eid = {A189},
        pages = {A189},
          doi = {10.1051/0004-6361/202452186},
archivePrefix = {arXiv},
       eprint = {2409.05948},
 primaryClass = {astro-ph.GA},
       adsurl = {https://ui.adsabs.harvard.edu/abs/2025A&A...697A.189D}
}

@ARTICLE{Lupton2004,
       author = {{Lupton}, Robert and {Blanton}, Michael R. and {Fekete}, George and {Hogg}, David W. and {O'Mullane}, Wil and {Szalay}, Alex and {Wherry}, Nicholas},
        title = "{Preparing Red-Green-Blue Images from CCD Data}",
      journal = {\pasp},
         year = 2004,
        month = feb,
       volume = {116},
       number = {816},
        pages = {133-137},
          doi = {10.1086/382245},
archivePrefix = {arXiv},
       eprint = {astro-ph/0312483},
 primaryClass = {astro-ph},
       adsurl = {https://ui.adsabs.harvard.edu/abs/2004PASP..116..133L}
}

@ARTICLE{vandenOord2018,
       author = {{van den Oord}, Aaron and {Li}, Yazhe and {Vinyals}, Oriol},
        title = "{Representation Learning with Contrastive Predictive Coding}",
      journal = {arXiv e-prints},
         year = 2018,
        month = jul,
          eid = {arXiv:1807.03748},
        pages = {arXiv:1807.03748},
          doi = {10.48550/arXiv.1807.03748},
archivePrefix = {arXiv},
       eprint = {1807.03748},
 primaryClass = {cs.LG},
       adsurl = {https://ui.adsabs.harvard.edu/abs/2018arXiv180703748V}
}

@ARTICLE{Lin2017,
       author = {{Lin}, Tsung-Yi and {Goyal}, Priya and {Girshick}, Ross and {He}, Kaiming and {Doll{\'a}r}, Piotr},
        title = "{Focal Loss for Dense Object Detection}",
      journal = {arXiv e-prints},
         year = 2017,
        month = aug,
          eid = {arXiv:1708.02002},
        pages = {arXiv:1708.02002},
          doi = {10.48550/arXiv.1708.02002},
archivePrefix = {arXiv},
       eprint = {1708.02002},
 primaryClass = {cs.CV},
       adsurl = {https://ui.adsabs.harvard.edu/abs/2017arXiv170802002L}
}

@INPROCEEDINGS{He2016,
       author = {{He}, Kaiming and {Zhang}, Xiangyu and {Ren}, Shaoqing and {Sun}, Jian},
        title = "{Deep Residual Learning for Image Recognition}",
    booktitle = {2016 IEEE Conference on Computer Vision and Pattern Recognition Workshops (CVPRW)},
         year = 2016,
        month = jun,
    publisher = {IEEE},
          eid = {1},
        pages = {1},
          doi = {10.1109/CVPR.2016.90},
archivePrefix = {arXiv},
       eprint = {1512.03385},
 primaryClass = {cs.CV},
       adsurl = {https://ui.adsabs.harvard.edu/abs/2016cvpr.confE...1H}
}

@ARTICLE{Dosovitskiy2020,
       author = {{Dosovitskiy}, Alexey and {Beyer}, Lucas and {Kolesnikov}, Alexander and {Weissenborn}, Dirk and {Zhai}, Xiaohua and {Unterthiner}, Thomas and {Dehghani}, Mostafa and {Minderer}, Matthias and {Heigold}, Georg and {Gelly}, Sylvain and {Uszkoreit}, Jakob and {Houlsby}, Neil},
        title = "{An Image is Worth 16x16 Words: Transformers for Image Recognition at Scale}",
      journal = {arXiv e-prints},
         year = 2020,
        month = oct,
          eid = {arXiv:2010.11929},
        pages = {arXiv:2010.11929},
          doi = {10.48550/arXiv.2010.11929},
archivePrefix = {arXiv},
       eprint = {2010.11929},
 primaryClass = {cs.CV},
       adsurl = {https://ui.adsabs.harvard.edu/abs/2020arXiv201011929D}
}

@ARTICLE{Vaswani2017,
       author = {{Vaswani}, Ashish and {Shazeer}, Noam and {Parmar}, Niki and {Uszkoreit}, Jakob and {Jones}, Llion and {Gomez}, Aidan N. and {Kaiser}, Lukasz and {Polosukhin}, Illia},
        title = "{Attention Is All You Need}",
      journal = {arXiv e-prints},
         year = 2017,
        month = jun,
          eid = {arXiv:1706.03762},
        pages = {arXiv:1706.03762},
          doi = {10.48550/arXiv.1706.03762},
archivePrefix = {arXiv},
       eprint = {1706.03762},
 primaryClass = {cs.CL},
       adsurl = {https://ui.adsabs.harvard.edu/abs/2017arXiv170603762V}
}

@ARTICLE{Loshchilov2019,
       author = {{Loshchilov}, Ilya and {Hutter}, Frank},
        title = "{Decoupled Weight Decay Regularization}",
      journal = {arXiv e-prints},
         year = 2017,
        month = nov,
          eid = {arXiv:1711.05101},
        pages = {arXiv:1711.05101},
          doi = {10.48550/arXiv.1711.05101},
archivePrefix = {arXiv},
       eprint = {1711.05101},
 primaryClass = {cs.LG},
       adsurl = {https://ui.adsabs.harvard.edu/abs/2017arXiv171105101L}
}

@ARTICLE{Sneppen2026,
       author = {{Sneppen}, A. and {Watson}, D. and {Matthews}, J.~H. and {Nikopoulos}, G. and {Allen}, N. and {Brammer}, G. and {Damgaard}, R. and {Heintz}, K.~E. and {Knigge}, C. and {Long}, K.~S. and {Rusakov}, V. and {Sim}, S.~A. and {Witstok}, J.},
        title = "{Inside the cocoon: a comprehensive explanation of the spectra of Little Red Dots}",
      journal = {arXiv e-prints},
         year = 2026,
        month = jan,
          eid = {arXiv:2601.18864},
        pages = {arXiv:2601.18864},
          doi = {10.48550/arXiv.2601.18864},
archivePrefix = {arXiv},
       eprint = {2601.18864},
 primaryClass = {astro-ph.GA},
       adsurl = {https://ui.adsabs.harvard.edu/abs/2026arXiv260118864S}
}

@article{McInnes2018,
       author = {{McInnes}, Leland and {Healy}, John and {Melville}, James},
        title = {UMAP: Uniform Manifold Approximation and Projection for Dimension Reduction},
      journal = {arXiv e-prints},
         year = {2018},
          eid = {arXiv:1802.03426},
        pages = {arXiv:1802.03426},
          doi = {10.48550/arXiv.1802.03426},
archivePrefix = {arXiv},
       eprint = {1802.03426},
 primaryClass = {stat.ML},
       adsurl = {https://ui.adsabs.harvard.edu/abs/2018arXiv180203426M}
}

@ARTICLE{Matthee2024,
       author = {{Matthee}, Jorryt and {Naidu}, Rohan P. and {Brammer}, Gabriel and {Chisholm}, John and {Eilers}, Anna-Christina and {Goulding}, Andy and {Greene}, Jenny and {Kashino}, Daichi and {Labbe}, Ivo and {Lilly}, Simon J. and {Mackenzie}, Ruari and {Oesch}, Pascal A. and {Weibel}, Andrea and {Wuyts}, Stijn and {Xiao}, Mengyuan and {Bordoloi}, Rongmon and {Bouwens}, Rychard and {van Dokkum}, Pieter and {Illingworth}, Garth and {Kramarenko}, Ivan and {Maseda}, Michael V. and {Mason}, Charlotte and {Meyer}, Romain A. and {Nelson}, Erica J. and {Reddy}, Naveen A. and {Shivaei}, Irene and {Simcoe}, Robert A. and {Yue}, Minghao},
        title = "{Little Red Dots: An Abundant Population of Faint Active Galactic Nuclei at z {\ensuremath{\sim}} 5 Revealed by the EIGER and FRESCO JWST Surveys}",
      journal = {\apj},
         year = 2024,
        month = mar,
       volume = {963},
       number = {2},
          eid = {129},
        pages = {129},
          doi = {10.3847/1538-4357/ad2345},
archivePrefix = {arXiv},
       eprint = {2306.05448},
 primaryClass = {astro-ph.GA},
       adsurl = {https://ui.adsabs.harvard.edu/abs/2024ApJ...963..129M}
}

@ARTICLE{Greene2024,
       author = {{Greene}, Jenny E. and {Labbe}, Ivo and {Goulding}, Andy D. and {Furtak}, Lukas J. and {Chemerynska}, Iryna and {Kokorev}, Vasily and {Dayal}, Pratika and {Volonteri}, Marta and {Williams}, Christina C. and {Wang}, Bingjie and {Setton}, David J. and {Burgasser}, Adam J. and {Bezanson}, Rachel and {Atek}, Hakim and {Brammer}, Gabriel and {Cutler}, Sam E. and {Feldmann}, Robert and {Fujimoto}, Seiji and {Glazebrook}, Karl and {de Graaff}, Anna and {Khullar}, Gourav and {Leja}, Joel and {Marchesini}, Danilo and {Maseda}, Michael V. and {Matthee}, Jorryt and {Miller}, Tim B. and {Naidu}, Rohan P. and {Nanayakkara}, Themiya and {Oesch}, Pascal A. and {Pan}, Richard and {Papovich}, Casey and {Price}, Sedona H. and {van Dokkum}, Pieter and {Weaver}, John R. and {Whitaker}, Katherine E. and {Zitrin}, Adi},
        title = "{UNCOVER Spectroscopy Confirms the Surprising Ubiquity of Active Galactic Nuclei in Red Sources at z > 5}",
      journal = {\apj},
         year = 2024,
        month = mar,
       volume = {964},
       number = {1},
          eid = {39},
        pages = {39},
          doi = {10.3847/1538-4357/ad1e5f},
archivePrefix = {arXiv},
       eprint = {2309.05714},
 primaryClass = {astro-ph.GA},
       adsurl = {https://ui.adsabs.harvard.edu/abs/2024ApJ...964...39G}
}

@ARTICLE{Labbe2025,
       author = {{Labbe}, Ivo and {Greene}, Jenny E. and {Bezanson}, Rachel and {Fujimoto}, Seiji and {Furtak}, Lukas J. and {Goulding}, Andy D. and {Matthee}, Jorryt and {Naidu}, Rohan P. and {Oesch}, Pascal A. and {Atek}, Hakim and {Brammer}, Gabriel and {Chemerynska}, Iryna and {Coe}, Dan and {Cutler}, Sam E. and {Dayal}, Pratika and {Feldmann}, Robert and {Franx}, Marijn and {Glazebrook}, Karl and {Leja}, Joel and {Maseda}, Michael and {Marchesini}, Danilo and {Nanayakkara}, Themiya and {Nelson}, Erica J. and {Pan}, Richard and {Papovich}, Casey and {Price}, Sedona H. and {Suess}, Katherine A. and {Wang}, Bingjie and {Weaver}, John R. and {Whitaker}, Katherine E. and {Williams}, Christina C. and {Zitrin}, Adi},
        title = "{UNCOVER: Candidate Red Active Galactic Nuclei at 3 < z < 7 with JWST and ALMA}",
      journal = {\apj},
         year = 2025,
        month = jan,
       volume = {978},
       number = {1},
          eid = {92},
        pages = {92},
          doi = {10.3847/1538-4357/ad3551},
archivePrefix = {arXiv},
       eprint = {2306.07320},
 primaryClass = {astro-ph.GA},
       adsurl = {https://ui.adsabs.harvard.edu/abs/2025ApJ...978...92L}
}

@ARTICLE{Akins2025,
       author = {{Akins}, Hollis B. and {Casey}, Caitlin M. and {Lambrides}, Erini and {Allen}, Natalie and {Andika}, Irham T. and {Brinch}, Malte and {Champagne}, Jaclyn B. and {Cooper}, Olivia and {Ding}, Xuheng and {Drakos}, Nicole E. and {Faisst}, Andreas and {Finkelstein}, Steven L. and {Franco}, Maximilien and {Fujimoto}, Seiji and {Gentile}, Fabrizio and {Gillman}, Steven and {Gozaliasl}, Ghassem and {Harish}, Santosh and {Hayward}, Christopher C. and {Hirschmann}, Michaela and {Ilbert}, Olivier and {Kartaltepe}, Jeyhan S. and {Kocevski}, Dale D. and {Koekemoer}, Anton M. and {Kokorev}, Vasily and {Liu}, Daizhong and {Long}, Arianna S. and {McCracken}, Henry Joy and {McKinney}, Jed and {Onoue}, Masafusa and {Paquereau}, Louise and {Renzini}, Alvio and {Rhodes}, Jason and {Robertson}, Brant E. and {Shuntov}, Marko and {Silverman}, John D. and {Tanaka}, Takumi S. and {Toft}, Sune and {Trakhtenbrot}, Benny and {Valentino}, Francesco and {Zavala}, Jorge},
        title = "{COSMOS-Web: The Overabundance and Physical Nature of ``Little Red Dots''{\textemdash}Implications for Early Galaxy and SMBH Assembly}",
      journal = {\apj},
         year = 2025,
        month = sep,
       volume = {991},
       number = {1},
          eid = {37},
        pages = {37},
          doi = {10.3847/1538-4357/ade984},
archivePrefix = {arXiv},
       eprint = {2406.10341},
 primaryClass = {astro-ph.GA},
       adsurl = {https://ui.adsabs.harvard.edu/abs/2025ApJ...991...37A}
}

@ARTICLE{Inayoshi&Luis2025,
       author = {{Inayoshi}, Kohei and {Ho}, Luis C.},
        title = "{A Critical Evaluation of the Physical Nature of the Little Red Dots}",
      journal = {arXiv e-prints},
         year = 2025,
        month = dec,
          eid = {arXiv:2512.03130},
        pages = {arXiv:2512.03130},
          doi = {10.48550/arXiv.2512.03130},
archivePrefix = {arXiv},
       eprint = {2512.03130},
 primaryClass = {astro-ph.GA},
       adsurl = {https://ui.adsabs.harvard.edu/abs/2025arXiv251203130I}
}

@ARTICLE{Chenchanghao2026,
       author = {{Chen}, Chang-Hao and {Shangguan}, Jinyi and {Ho}, Luis C. and {Zhang}, Zijian and {Inayoshi}, Kohei and {Li}, Ruancun},
        title = "{ABCD: The Nuclear Structure of the Little Red Dots Revealted through Absorption, Break, Continuum, and Decrement}",
      journal = {arXiv e-prints},
         year = 2026,
        month = jun,
          eid = {arXiv:2606.04711},
        pages = {arXiv:2606.04711},
          doi = {10.48550/arXiv.2606.04711},
archivePrefix = {arXiv},
       eprint = {2606.04711},
 primaryClass = {astro-ph.GA},
       adsurl = {https://ui.adsabs.harvard.edu/abs/2026arXiv260604711C}
}

@ARTICLE{Inayoshi&Maiolino2025,
       author = {{Inayoshi}, Kohei and {Maiolino}, Roberto},
        title = "{Extremely Dense Gas around Little Red Dots and High-redshift Active Galactic Nuclei: A Nonstellar Origin of the Balmer Break and Absorption Features}",
      journal = {\apjl},
         year = 2025,
        month = feb,
       volume = {980},
       number = {2},
          eid = {L27},
        pages = {L27},
          doi = {10.3847/2041-8213/adaebd},
archivePrefix = {arXiv},
       eprint = {2409.07805},
 primaryClass = {astro-ph.GA},
       adsurl = {https://ui.adsabs.harvard.edu/abs/2025ApJ...980L..27I}
}

@ARTICLE{Rusakov2026,
       author = {{Rusakov}, V. and {Watson}, D. and {Nikopoulos}, G.~P. and {Brammer}, G. and {Gottumukkala}, R. and {Harvey}, T. and {Heintz}, K.~E. and {Damgaard}, R. and {Sim}, S.~A. and {Sneppen}, A. and {Vijayan}, A.~P. and {Adams}, N. and {Austin}, D. and {Conselice}, C.~J. and {Goolsby}, C.~M. and {Toft}, S. and {Witstok}, J.},
        title = "{Little red dots as young supermassive black holes in dense ionized cocoons}",
      journal = {\nat},
         year = 2026,
        month = jan,
       volume = {649},
       number = {8097},
        pages = {574-579},
          doi = {10.1038/s41586-025-09900-4},
archivePrefix = {arXiv},
       eprint = {2503.16595},
 primaryClass = {astro-ph.GA},
       adsurl = {https://ui.adsabs.harvard.edu/abs/2026Natur.649..574R}
}

@ARTICLE{Naidu2025,
       author = {{Naidu}, Rohan P. and {Matthee}, Jorryt and {Katz}, Harley and {de Graaff}, Anna and {Oesch}, Pascal and {Smith}, Aaron and {Greene}, Jenny E. and {Brammer}, Gabriel and {Weibel}, Andrea and {Hviding}, Raphael and {Chisholm}, John and {Labb\textbackslash'e}, Ivo and {Simcoe}, Robert A. and {Witten}, Callum and {Atek}, Hakim and {Baggen}, Josephine F.~W. and {Belli}, Sirio and {Bezanson}, Rachel and {Boogaard}, Leindert A. and {Bose}, Sownak and {Covelo-Paz}, Alba and {Dayal}, Pratika and {Fudamoto}, Yoshinobu and {Furtak}, Lukas J. and {Giovinazzo}, Emma and {Goulding}, Andy and {Gronke}, Max and {Heintz}, Kasper E. and {Hirschmann}, Michaela and {Illingworth}, Garth and {Inoue}, Akio K. and {Johnson}, Benjamin D. and {Leja}, Joel and {Leonova}, Ecaterina and {McConachie}, Ian and {Maseda}, Michael V. and {Natarajan}, Priyamvada and {Nelson}, Erica and {Setton}, David J. and {Shivaei}, Irene and {Sobral}, David and {Stefanon}, Mauro and {Tacchella}, Sandro and {Toft}, Sune and {Torralba}, Alberto and {van Dokkum}, Pieter and {van der Wel}, Arjen and {Volonteri}, Marta and {Walter}, Fabian and {Wang}, Bingjie and {Watson}, Darach},
        title = "{A ``Black Hole Star'' Reveals the Remarkable Gas-Enshrouded Hearts of the Little Red Dots}",
      journal = {arXiv e-prints},
         year = 2025,
        month = mar,
          eid = {arXiv:2503.16596},
        pages = {arXiv:2503.16596},
          doi = {10.48550/arXiv.2503.16596},
archivePrefix = {arXiv},
       eprint = {2503.16596},
 primaryClass = {astro-ph.GA},
       adsurl = {https://ui.adsabs.harvard.edu/abs/2025arXiv250316596N}
}

@ARTICLE{PerezGonzalez2024,
       author = {{P{\'e}rez-Gonz{\'a}lez}, Pablo G. and {Barro}, Guillermo and {Rieke}, George H. and {Lyu}, Jianwei and {Rieke}, Marcia and {Alberts}, Stacey and {Williams}, Christina C. and {Hainline}, Kevin and {Sun}, Fengwu and {Pusk{\'a}s}, D{\'a}vid and {Annunziatella}, Marianna and {Baker}, William M. and {Bunker}, Andrew J. and {Egami}, Eiichi and {Ji}, Zhiyuan and {Johnson}, Benjamin D. and {Robertson}, Brant and {Rodr{\'\i}guez Del Pino}, Bruno and {Rujopakarn}, Wiphu and {Shivaei}, Irene and {Tacchella}, Sandro and {Willmer}, Christopher N.~A. and {Willott}, Chris},
        title = "{What Is the Nature of Little Red Dots and what Is Not, MIRI SMILES Edition}",
      journal = {\apj},
         year = 2024,
        month = jun,
       volume = {968},
       number = {1},
          eid = {4},
        pages = {4},
          doi = {10.3847/1538-4357/ad38bb},
archivePrefix = {arXiv},
       eprint = {2401.08782},
 primaryClass = {astro-ph.GA},
       adsurl = {https://ui.adsabs.harvard.edu/abs/2024ApJ...968....4P}
}

@ARTICLE{Dieleman2015,
       author = {{Dieleman}, Sander and {Willett}, Kyle W. and {Dambre}, Joni},
        title = "{Rotation-invariant convolutional neural networks for galaxy morphology prediction}",
      journal = {\mnras},
         year = 2015,
        month = jun,
       volume = {450},
       number = {2},
        pages = {1441-1459},
          doi = {10.1093/mnras/stv632},
archivePrefix = {arXiv},
       eprint = {1503.07077},
 primaryClass = {astro-ph.IM},
       adsurl = {https://ui.adsabs.harvard.edu/abs/2015MNRAS.450.1441D}
}

@ARTICLE{Walmsley2022,
       author = {{Walmsley}, Mike and {Lintott}, Chris and {G{\'e}ron}, Tobias and {Kruk}, Sandor and {Krawczyk}, Coleman and {Willett}, Kyle W. and {Bamford}, Steven and {Kelvin}, Lee S. and {Fortson}, Lucy and {Gal}, Yarin and {Keel}, William and {Masters}, Karen L. and {Mehta}, Vihang and {Simmons}, Brooke D. and {Smethurst}, Rebecca and {Smith}, Lewis and {Baeten}, Elisabeth M. and {Macmillan}, Christine},
        title = "{Galaxy Zoo DECaLS: Detailed visual morphology measurements from volunteers and deep learning for 314 000 galaxies}",
      journal = {\mnras},
         year = 2022,
        month = jan,
       volume = {509},
       number = {3},
        pages = {3966-3988},
          doi = {10.1093/mnras/stab2093},
archivePrefix = {arXiv},
       eprint = {2102.08414},
 primaryClass = {astro-ph.GA},
       adsurl = {https://ui.adsabs.harvard.edu/abs/2022MNRAS.509.3966W}
}

@ARTICLE{Pasquet2019,
       author = {{Pasquet}, Johanna and {Bertin}, E. and {Treyer}, M. and {Arnouts}, S. and {Fouchez}, D.},
        title = "{Photometric redshifts from SDSS images using a convolutional neural network}",
      journal = {\aap},
         year = 2019,
        month = jan,
       volume = {621},
          eid = {A26},
        pages = {A26},
          doi = {10.1051/0004-6361/201833617},
archivePrefix = {arXiv},
       eprint = {1806.06607},
 primaryClass = {astro-ph.IM},
       adsurl = {https://ui.adsabs.harvard.edu/abs/2019A&A...621A..26P}
}

@ARTICLE{Parker2024,
       author = {{Parker}, Liam and {Lanusse}, Francois and {Golkar}, Siavash and {Sarra}, Leopoldo and {Cranmer}, Miles and {Bietti}, Alberto and {Eickenberg}, Michael and {Krawezik}, Geraud and {McCabe}, Michael and {Morel}, Rudy and {Ohana}, Ruben and {Pettee}, Mariel and {R{\'e}galdo-Saint Blancard}, Bruno and {Cho}, Kyunghyun and {Ho}, Shirley and {Polymathic AI Collaboration}},
        title = "{AstroCLIP: a cross-modal foundation model for galaxies}",
      journal = {\mnras},
         year = 2024,
        month = jul,
       volume = {531},
       number = {4},
        pages = {4990-5011},
          doi = {10.1093/mnras/stae1450},
archivePrefix = {arXiv},
       eprint = {2310.03024},
 primaryClass = {astro-ph.IM},
       adsurl = {https://ui.adsabs.harvard.edu/abs/2024MNRAS.531.4990P}
}

@article{Meng2026,
doi = {10.3847/2041-8213/aea06f},
url = {https://doi.org/10.3847/2041-8213/aea06f},
year = {2026},
month = {oct},
publisher = {The American Astronomical Society},
volume = {1009},
number = {2},
pages = {L56},
author = {Meng, Hao and Ye, Guangping and Zhang, Huanian},
title = {Probing Dark Matter Halos of High-redshift Quasars via Wide-field Clustering},
journal = {The Astrophysical Journal Letters},
}

@ARTICLE{Barro2026,
       author = {{Barro}, Guillermo and {P{\'e}rez-Gonz{\'a}lez}, Pablo G. and {Kocevski}, Dale and {Trump}, Jonathan R. and {Dickinson}, Mark and {Arrabal Haro}, Pablo and {Brooks}, Madisyn and {Donnan}, Callum T. and {Dunlop}, James S. and {Finkelstein}, Steven L. and {Franco}, Maximilien and {Gandolfi}, Giovanni and {Giavalisco}, Mauro and {Grogin}, Norman A. and {Hirschmann}, Michaela and {Kartaltepe}, Jeyhan S. and {Koekemoer}, Anton M. and {Larson}, Rebecca L. and {Leung}, Gene C.~K. and {Lucas}, Ray A. and {McGrath}, Elizabeth J. and {Papovich}, Casey and {P{\'e}rez-D{\'\i}az}, Borja and {Somerville}, Rachel S. and {Taylor}, Elizabeth and {Taylor}, Anthony J. and {Tripodi}, Roberta and {Yung}, L.~Y. Aaron and {Wang}, Xin},
        title = "{From ``The Cliff'' to ``Virgil'': Mapping the Spectral Diversity of Little Red Dots with JWST/NIRSpec}",
      journal = {\apj},
         year = 2026,
        month = may,
       volume = {1003},
       number = {1},
          eid = {96},
        pages = {96},
          doi = {10.3847/1538-4357/ae5d2c},
archivePrefix = {arXiv},
       eprint = {2512.15853},
 primaryClass = {astro-ph.GA},
       adsurl = {https://ui.adsabs.harvard.edu/abs/2026ApJ..1003...96B}
}

@ARTICLE{Barro2024,
       author = {{Barro}, Guillermo and {P{\'e}rez-Gonz{\'a}lez}, Pablo G. and {Kocevski}, Dale D. and {McGrath}, Elizabeth J. and {Trump}, Jonathan R. and {Simons}, Raymond C. and {Somerville}, Rachel S. and {Yung}, L.~Y. Aaron and {Arrabal Haro}, Pablo and {Akins}, Hollis B. and {Bagley}, Michaela B. and {Cleri}, Nikko J. and {Costantin}, Luca and {Davis}, Kelcey and {Dickinson}, Mark and {Finkelstein}, Steve L. and {Giavalisco}, Mauro and {G{\'o}mez-Guijarro}, Carlos and {Hathi}, Nimish P. and {Hirschmann}, Michaela and {Holwerda}, Benne W. and {Huertas-Company}, Marc and {Kartaltepe}, Jeyhan S. and {Koekemoer}, Anton M. and {Lucas}, Ray A. and {Papovich}, Casey and {Pirzkal}, Nor and {Seill{\'e}}, Lise-Marie and {Tacchella}, Sandro and {Wuyts}, Stijn and {Wilkins}, Stephen M. and {de la Vega}, Alexander and {Yang}, Guang and {Zavala}, Jorge A.},
        title = "{Extremely Red Galaxies at z = 5─9 with MIRI and NIRSpec: Dusty Galaxies or Obscured Active Galactic Nuclei?}",
      journal = {\apj},
         year = 2024,
        month = mar,
       volume = {963},
       number = {2},
          eid = {128},
        pages = {128},
          doi = {10.3847/1538-4357/ad167e},
archivePrefix = {arXiv},
       eprint = {2305.14418},
 primaryClass = {astro-ph.GA},
       adsurl = {https://ui.adsabs.harvard.edu/abs/2024ApJ...963..128B}
}

@ARTICLE{Sun2025,
       author = {{Sun}, Fengwu and {Fudamoto}, Yoshinobu and {Lin}, Xiaojing and {Helton}, Jakob M. and {Hsiao}, Tiger Yu-Yang and {Egami}, Eiichi and {Akhtarkavan}, Arshia and {Bunker}, Andrew J. and {Cai}, Zheng and {DeCoursey}, Christa and {Eisenstein}, Daniel J. and {Fan}, Xiaohui and {Harikane}, Yuichi and {Ji}, Zhiyuan and {Jin}, Xiangyu and {Liu}, Weizhe and {Liu}, Yichen and {Ma}, Zheng and {Maiolino}, Roberto and {Ouchi}, Masami and {Tee}, Wei Leong and {Wang}, Feige and {Willmer}, Christopher N.~A. and {Wu}, Yunjing and {Xu}, Yi and {Yang}, Jinyi and {Zhang}, Junyu and {Zhu}, Yongda},
        title = "{Slitless Areal Pure-Parallel HIgh-Redshift Emission Survey (SAPPHIRES): Early Data Release of Deep JWST/NIRCam Images and Spectra in MACS J0416 Parallel Field}",
      journal = {arXiv e-prints},
         year = 2025,
        month = mar,
          eid = {arXiv:2503.15587},
        pages = {arXiv:2503.15587},
          doi = {10.48550/arXiv.2503.15587},
archivePrefix = {arXiv},
       eprint = {2503.15587},
 primaryClass = {astro-ph.GA},
       adsurl = {https://ui.adsabs.harvard.edu/abs/2025arXiv250315587S}
}

@ARTICLE{Morishita2025,
       author = {{Morishita}, Takahiro and {Mason}, Charlotte A. and {Kreilgaard}, Kimi C. and {Trenti}, Michele and {Treu}, Tommaso and {Vulcani}, Benedetta and {Zhang}, Yechi and {Abdurro'uf} and {Alavi}, Anahita and {Atek}, Hakim and {Bah{\'e}}, Yannick and {Brada{\v{c}}}, Maru{\v{s}}a and {Bradley}, Larry D. and {Bunker}, Andrew J. and {Coe}, Dan and {Colbert}, James and {Gelli}, Viola and {Hayes}, Matthew J. and {Jones}, Tucker and {Kodama}, Tadayuki and {Leethochawalit}, Nicha and {Liu}, Zhaoran and {Malkan}, Matthew A. and {Mehta}, Vihang and {Metha}, Benjamin and {Newman}, Andrew B. and {Rafelski}, Marc and {Roberts-Borsani}, Guido and {Rutkowski}, Michael J. and {Scarlata}, Claudia and {Stiavelli}, Massimo and {Sutanto}, Ryo A. and {Takahashi}, Kosuke and {Teplitz}, Harry I. and {Wang}, Xin},
        title = "{BEACON: JWST NIRCam Pure-parallel Imaging Survey. I. Survey Design and Initial Results}",
      journal = {\apj},
         year = 2025,
        month = apr,
       volume = {983},
       number = {2},
          eid = {152},
        pages = {152},
          doi = {10.3847/1538-4357/adbbdc},
archivePrefix = {arXiv},
       eprint = {2412.04211},
 primaryClass = {astro-ph.GA},
       adsurl = {https://ui.adsabs.harvard.edu/abs/2025ApJ...983..152M}
}

@ARTICLE{Rinaldi2026,
       author = {{Rinaldi}, Pierluigi and {Hainline}, Kevin and {D'Eugenio}, Francesco and {P{\'e}rez-Gonz{\'a}lez}, Pablo G. and {Eisenstein}, Daniel J. and {Willmer}, Christopher N.~A. and {Carreira}, Courtney and {Robertson}, Brant and {Johnson}, Benjamin D. and {Alberts}, Stacey and {Baker}, William M. and {Bunker}, Andrew J. and {Carniani}, Stefano and {Egami}, Eiichi and {Helton}, Jakob M. and {Ji}, Zhiyuan and {Juod{\v{z}}balis}, Ignas and {Lin}, Xiaojing and {Lyu}, Jianwei and {Ma}, Zheng and {Maiolino}, Roberto and {Parlanti}, Eleonora and {Scholtz}, Jan and {Sun}, Yang and {Tacchella}, Sandro and {Venturi}, Giacomo and {Williams}, Christina C. and {Willott}, Chris and {Witstok}, Joris and {Wu}, Zihao},
        title = "{The Way We Tally Becomes the Tale: the Impact of Selection Strategies on the Inferred Evolution of Little Red Dots Across Cosmic Time}",
      journal = {arXiv e-prints},
         year = 2026,
        month = apr,
          eid = {arXiv:2604.07138},
        pages = {arXiv:2604.07138},
          doi = {10.48550/arXiv.2604.07138},
archivePrefix = {arXiv},
       eprint = {2604.07138},
 primaryClass = {astro-ph.GA},
       adsurl = {https://ui.adsabs.harvard.edu/abs/2026arXiv260407138R}
}

@ARTICLE{Heintz2024DJA,
       author = {{Heintz}, Kasper E. and {Watson}, Darach and {Brammer}, Gabriel and {Vejlgaard}, Simone and {Hutter}, Anne and {Strait}, Victoria B. and {Matthee}, Jorryt and {Oesch}, Pascal A. and {Jakobsson}, P{\'a}ll and {Tanvir}, Nial R. and {Laursen}, Peter and {Naidu}, Rohan P. and {Mason}, Charlotte A. and {Killi}, Meghana and {Jung}, Intae and {Hsiao}, Tiger Yu-Yang and {Abdurro'uf} and {Coe}, Dan and {Arrabal Haro}, Pablo and {Finkelstein}, Steven L. and {Toft}, Sune},
        title = "{Strong damped Lyman-{\ensuremath{\alpha}} absorption in young star-forming galaxies at redshifts 9 to 11}",
      journal = {Science},
         year = 2024,
        month = may,
       volume = {384},
       number = {6698},
        pages = {890-894},
          doi = {10.1126/science.adj0343},
archivePrefix = {arXiv},
       eprint = {2306.00647},
 primaryClass = {astro-ph.GA},
       adsurl = {https://ui.adsabs.harvard.edu/abs/2024Sci...384..890H}
}

@ARTICLE{Radford2021,
       author = {{Radford}, Alec and {Kim}, Jong Wook and {Hallacy}, Chris and {Ramesh}, Aditya and {Goh}, Gabriel and {Agarwal}, Sandhini and {Sastry}, Girish and {Askell}, Amanda and {Mishkin}, Pamela and {Clark}, Jack and {Krueger}, Gretchen and {Sutskever}, Ilya},
        title = "{Learning Transferable Visual Models From Natural Language Supervision}",
      journal = {arXiv e-prints},
         year = 2021,
        month = feb,
          eid = {arXiv:2103.00020},
        pages = {arXiv:2103.00020},
          doi = {10.48550/arXiv.2103.00020},
archivePrefix = {arXiv},
       eprint = {2103.00020},
 primaryClass = {cs.CV},
       adsurl = {https://ui.adsabs.harvard.edu/abs/2021arXiv210300020R}
}

@ARTICLE{Brazzini2026,
       author = {{Brazzini}, M. and {D'Eugenio}, F. and {Maiolino}, R. and {Lyu}, J. and {DeCoursey}, C. and {{\"U}bler}, H. and {Ji}, X. and {Juod{\v{z}}balis}, I. and {Scholtz}, J. and {Jones}, G.~C. and {Hainline}, K. and {Dalla Bont{\`a}}, E. and {{\'e}rez-Gonz{\'a}lez}, P.~G. P and {Geris}, S. and {Harshan}, A. and {Feruglio}, C. and {Bischetti}, M. and {Mazzolari}, G. and {Rieke}, G. and {Alberts}, S. and {Trefoloni}, B. and {Carniani}, S. and {Parlanti}, E. and {Marconi}, A. and {Risaliti}, G. and {Ramos Almeida}, C. and {Rinaldi}, P. and {Perna}, M. and {Zamora}, S. and {Lamperti}, I. and {Venturi}, G. and {Cresci}, G. and {Bunker}, Andrew J. and {Ivey}, L.~R.},
        title = "{The Little Blue and Red Dots Rosetta Stones: Non-Gaussian broad lines, hot dust, and X-ray weakness}",
      journal = {arXiv e-prints},
         year = 2026,
        month = jan,
          eid = {arXiv:2601.22214},
        pages = {arXiv:2601.22214},
          doi = {10.48550/arXiv.2601.22214},
archivePrefix = {arXiv},
       eprint = {2601.22214},
 primaryClass = {astro-ph.GA},
       adsurl = {https://ui.adsabs.harvard.edu/abs/2026arXiv260122214B}
}

@ARTICLE{Brazzini2025,
       author = {{Brazzini}, Matilde and {D'Eugenio}, Francesco and {Maiolino}, Roberto and {Juod{\v{z}}balis}, Ignas and {Ji}, Xihan and {Scholtz}, Jan and {Chang}, Seok-Jun},
        title = "{Ruling out dominant electron scattering in Little Red Dots' Rosetta Stone using multiple hydrogen lines}",
      journal = {\mnras},
         year = 2025,
        month = nov,
       volume = {544},
       number = {1},
        pages = {L167-L173},
          doi = {10.1093/mnrasl/slaf116},
archivePrefix = {arXiv},
       eprint = {2507.08929},
 primaryClass = {astro-ph.GA},
       adsurl = {https://ui.adsabs.harvard.edu/abs/2025MNRAS.544L.167B}
}

@ARTICLE{Ginolfi2026,
       author = {{Ginolfi}, Michele and {Mannucci}, Filippo and {Marconi}, Alessandro and {D'Eugenio}, Francesco and {Venturi}, Giacomo and {Belfiore}, Francesco and {Cresci}, Giovanni and {Bracci}, Caterina and {Carniani}, Stefano and {Cozzi}, Alessandra and {Maiolino}, Roberto and {Risaliti}, Guido},
        title = "{Unsupervised selection and characterisation of Little Red Dots in JWST surveys with manifold learning}",
      journal = {arXiv e-prints},
         year = 2026,
        month = jul,
          eid = {arXiv:2607.22835},
        pages = {arXiv:2607.22835},
archivePrefix = {arXiv},
       eprint = {2607.22835},
 primaryClass = {astro-ph.GA},
       adsurl = {https://ui.adsabs.harvard.edu/abs/2026arXiv260722835G}
}

@ARTICLE{Merz2026,
       author = {{Merz}, Grant and {Zhuang}, Ming-Yang and {Li}, Junyao and {Yang}, Qian and {Shen}, Yue and {Liu}, Xin and {Crenshaw}, John Franklin},
        title = "{Photometric Redshifts in JWST Deep Fields: A Pixel-Based Alternative with DeepDISC}",
      journal = {The Open Journal of Astrophysics},
         year = 2026,
        month = feb,
       volume = {9},
        pages = {56099},
          doi = {10.33232/001c.156099},
archivePrefix = {arXiv},
       eprint = {2510.27032},
 primaryClass = {astro-ph.IM},
       adsurl = {https://ui.adsabs.harvard.edu/abs/2026OJAp....956099M}
}

@ARTICLE{Fudamoto2025,
       author = {{Fudamoto}, Yoshinobu and {Helton}, Jakob M. and {Lin}, Xiaojing and {Sun}, Fengwu and {Behroozi}, Peter and {Hsiao}, Tiger Yu-Yang and {Egami}, Eiichi and {Bunker}, Andrew J. and {Harikane}, Yuichi and {Ouchi}, Masami and {Liu}, Yichen and {Liu}, Weizhe and {Maiolino}, Roberto and {Ji}, Zhiyuan and {Jin}, Xiangyu and {Tee}, Wei Leong and {Wang}, Feige and {Willmer}, Christopher N.~A. and {Xu}, Yi and {Zhu}, Yongda},
        title = "{SAPPHIRES: A Galaxy Over-Density in the Heart of Cosmic Reionization at $z=8.47$}",
      journal = {arXiv e-prints},
         year = 2025,
        month = mar,
          eid = {arXiv:2503.15597},
        pages = {arXiv:2503.15597},
          doi = {10.48550/arXiv.2503.15597},
archivePrefix = {arXiv},
       eprint = {2503.15597},
 primaryClass = {astro-ph.GA},
       adsurl = {https://ui.adsabs.harvard.edu/abs/2025arXiv250315597F}
}

@ARTICLE{deGraaff2025b,
       author = {{de Graaff}, Anna and {Rix}, Hans-Walter and {Naidu}, Rohan P. and {Labb{\'e}}, Ivo and {Wang}, Bingjie and {Leja}, Joel and {Matthee}, Jorryt and {Katz}, Harley and {Greene}, Jenny E. and {Hviding}, Raphael E. and {Baggen}, Josephine and {Bezanson}, Rachel and {Boogaard}, Leindert A. and {Brammer}, Gabriel and {Dayal}, Pratika and {van Dokkum}, Pieter and {Goulding}, Andy D. and {Hirschmann}, Michaela and {Maseda}, Michael V. and {McConachie}, Ian and {Miller}, Tim B. and {Nelson}, Erica and {Oesch}, Pascal A. and {Setton}, David J. and {Shivaei}, Irene and {Weibel}, Andrea and {Whitaker}, Katherine E. and {Williams}, Christina C.},
        title = "{A remarkable ruby: Absorption in dense gas, rather than evolved stars, drives the extreme Balmer break of a little red dot at z = 3.5}",
      journal = {\aap},
         year = 2025,
        month = sep,
       volume = {701},
          eid = {A168},
        pages = {A168},
          doi = {10.1051/0004-6361/202554681},
archivePrefix = {arXiv},
       eprint = {2503.16600},
 primaryClass = {astro-ph.GA},
       adsurl = {https://ui.adsabs.harvard.edu/abs/2025A&A...701A.168D}
}

@ARTICLE{Wangbingjie2025,
       author = {{Wang}, Bingjie and {de Graaff}, Anna and {Davies}, Rebecca L. and {Greene}, Jenny E. and {Leja}, Joel and {Brammer}, Gabriel B. and {Goulding}, Andy D. and {Miller}, Tim B. and {Suess}, Katherine A. and {Weibel}, Andrea and {Williams}, Christina C. and {Bezanson}, Rachel and {Boogaard}, Leindert A. and {Cleri}, Nikko J. and {Hirschmann}, Michaela and {Katz}, Harley and {Labb{\'e}}, Ivo and {Maseda}, Michael V. and {Matthee}, Jorryt and {McConachie}, Ian and {Naidu}, Rohan P. and {Oesch}, Pascal A. and {Rix}, Hans-Walter and {Setton}, David J. and {Whitaker}, Katherine E.},
        title = "{RUBIES: JWST/NIRSpec Confirmation of an Infrared-luminous, Broad-line Little Red Dot with an Ionized Outflow}",
      journal = {\apj},
         year = 2025,
        month = may,
       volume = {984},
       number = {2},
          eid = {121},
        pages = {121},
          doi = {10.3847/1538-4357/adc1ca},
archivePrefix = {arXiv},
       eprint = {2403.02304},
 primaryClass = {astro-ph.GA},
       adsurl = {https://ui.adsabs.harvard.edu/abs/2025ApJ...984..121W}
}

@ARTICLE{Zhangyunchong2026,
       author = {{Zhang}, Yunchong and {de Graaff}, Anna and {Setton}, David J. and {Price}, Sedona H. and {Bezanson}, Rachel and {del P. Lagos}, Claudia and {Cutler}, Sam E. and {McConachie}, Ian and {Cleri}, Nikko J. and {Cooper}, Olivia R. and {Gottumukkala}, Rashmi and {Greene}, Jenny E. and {Hirschmann}, Michaela and {Khullar}, Gourav and {Labbe}, Ivo and {Leja}, Joel and {Maseda}, Michael V. and {Matthee}, Jorryt and {Miller}, Tim B. and {Nanayakkara}, Themiya and {Suess}, Katherine A. and {Wang}, Bingjie and {Whitaker}, Katherine E. and {Williams}, Christina C.},
        title = "{RUBIES Spectroscopically Confirms the High Number Density of Quiescent Galaxies from 2 < z< 5}",
      journal = {\apj},
         year = 2026,
        month = feb,
       volume = {997},
       number = {2},
          eid = {252},
        pages = {252},
          doi = {10.3847/1538-4357/ae24e1},
archivePrefix = {arXiv},
       eprint = {2508.08577},
 primaryClass = {astro-ph.GA},
       adsurl = {https://ui.adsabs.harvard.edu/abs/2026ApJ...997..252Z}
}

@ARTICLE{Ito2026,
       author = {{Ito}, K. and {Valentino}, F. and {Brammer}, G. and {Hamadouche}, M.~L. and {Whitaker}, K.~E. and {Kokorev}, V. and {Zhu}, P. and {Kakimoto}, T. and {Wu}, P.-F. and {Antwi-Danso}, J. and {Baker}, W.~M. and {Ceverino}, D. and {Faisst}, A.~L. and {Farcy}, M. and {Fujimoto}, S. and {Gallazzi}, A. and {Gillman}, S. and {Gottumukkala}, R. and {Heintz}, K.~E. and {Hirschmann}, M. and {Jespersen}, C.~K. and {Kubo}, M. and {Lee}, M. and {Magdis}, G. and {Onodera}, M. and {Shimakawa}, R. and {Tanaka}, M. and {Toft}, S. and {Weaver}, J.~R.},
        title = "{DeepDive: A deep dive into the physics of the first massive quiescent galaxies in the Universe}",
      journal = {\aap},
         year = 2026,
        month = jun,
       volume = {710},
          eid = {A364},
        pages = {A364},
          doi = {10.1051/0004-6361/202556137},
archivePrefix = {arXiv},
       eprint = {2506.22642},
 primaryClass = {astro-ph.GA},
       adsurl = {https://ui.adsabs.harvard.edu/abs/2026A&A...710A.364I}
}

@ARTICLE{Baker2025,
       author = {{Baker}, William M. and {Lim}, Seunghwan and {D'Eugenio}, Francesco and {Maiolino}, Roberto and {Ji}, Zhiyuan and {Arribas}, Santiago and {Bunker}, Andrew J. and {Carniani}, Stefano and {Charlot}, Stephane and {de Graaff}, Anna and {Hainline}, Kevin and {Looser}, Tobias J. and {Lyu}, Jianwei and {Rinaldi}, Pierluigi and {Robertson}, Brant and {Schaller}, Matthieu and {Schaye}, Joop and {Scholtz}, Jan and {{\"U}bler}, Hannah and {Williams}, Christina C. and {Willmer}, Christopher N.~A. and {Willott}, Chris and {Zhu}, Yongda},
        title = "{The abundance and nature of high-redshift quiescent galaxies from JADES spectroscopy and the FLAMINGO simulations}",
      journal = {\mnras},
         year = 2025,
        month = may,
       volume = {539},
       number = {1},
        pages = {557-589},
          doi = {10.1093/mnras/staf475},
archivePrefix = {arXiv},
       eprint = {2410.14773},
 primaryClass = {astro-ph.GA},
       adsurl = {https://ui.adsabs.harvard.edu/abs/2025MNRAS.539..557B}
}

@ARTICLE{CoveloPaz2026,
       author = {{Covelo-Paz}, Alba and {Meuwly}, Corentin and {Oesch}, Pascal A. and {Witten}, Callum and {Weibel}, Andrea and {Carvajal-Bohorquez}, Cristian and {Ciesla}, Laure and {Giovinazzo}, Emma and {Brammer}, Gabriel},
        title = "{A systematic search for dormant galaxies at z {\ensuremath{\sim}} 5{\ensuremath{-}}7 from the JWST NIRSpec archive}",
      journal = {\aap},
         year = 2026,
        month = jan,
       volume = {705},
          eid = {A155},
        pages = {A155},
          doi = {10.1051/0004-6361/202556119},
archivePrefix = {arXiv},
       eprint = {2506.22540},
 primaryClass = {astro-ph.GA},
       adsurl = {https://ui.adsabs.harvard.edu/abs/2026A&A...705A.155C}
}

@ARTICLE{Bertin1996,
       author = {{Bertin}, E. and {Arnouts}, S.},
        title = "{SExtractor: Software for source extraction.}",
      journal = {\aaps},
         year = 1996,
        month = jun,
       volume = {117},
        pages = {393-404},
          doi = {10.1051/aas:1996164},
       adsurl = {https://ui.adsabs.harvard.edu/abs/1996A&AS..117..393B}
}

@ARTICLE{Chenyixian2026,
       author = {{Chen}, Yi-Xian and {Liu}, Hanpu and {Li}, Ruancun and {Wang}, Bingjie and {Ma}, Yilun and {Jiang}, Yan-Fei and {Greene}, Jenny E. and {Quataert}, Eliot and {Goodman}, Jeremy},
        title = "{Spectral Appearance of Self-gravitating Disks Powered by Stellar Objects: Universal Effective Temperature in the Optical Continuum and Application to Little Red Dots}",
      journal = {\apjl},
         year = 2026,
        month = jun,
       volume = {1003},
       number = {2},
          eid = {L46},
        pages = {L46},
          doi = {10.3847/2041-8213/ae6b69},
archivePrefix = {arXiv},
       eprint = {2602.06954},
 primaryClass = {astro-ph.HE},
       adsurl = {https://ui.adsabs.harvard.edu/abs/2026ApJ..1003L..46C}
}

@ARTICLE{Zwick2026,
       author = {{Zwick}, Lorenz and {Tiede}, Christopher and {Mayer}, Lucio},
        title = "{Little Red Dots as Self-gravitating Disks Accreting on Supermassive Stars: Spectral Appearance and Formation Pathway of the Progenitors to Direct Collapse Black Holes}",
      journal = {\apj},
         year = 2026,
        month = may,
       volume = {1002},
       number = {1},
          eid = {7},
        pages = {7},
          doi = {10.3847/1538-4357/ae4d47},
archivePrefix = {arXiv},
       eprint = {2507.22014},
 primaryClass = {astro-ph.GA},
       adsurl = {https://ui.adsabs.harvard.edu/abs/2026ApJ..1002....7Z}
}

@ARTICLE{Pacucci2025,
       author = {{Pacucci}, Fabio and {Loeb}, Abraham},
        title = "{Cosmic Outliers: Low-spin Halos Explain the Abundance, Compactness, and Redshift Evolution of the Little Red Dots}",
      journal = {\apjl},
         year = 2025,
        month = aug,
       volume = {989},
       number = {2},
          eid = {L19},
        pages = {L19},
          doi = {10.3847/2041-8213/ade871},
archivePrefix = {arXiv},
       eprint = {2506.03244},
 primaryClass = {astro-ph.GA},
       adsurl = {https://ui.adsabs.harvard.edu/abs/2025ApJ...989L..19P}
}

@ARTICLE{Sunwendy2026,
       author = {{Sun}, Wendy Q. and {Naidu}, Rohan P. and {Matthee}, Jorryt and {de Graaff}, Anna and {Chisholm}, John and {Greene}, Jenny E. and {Oesch}, Pascal A. and {Torralba}, Alberto and {Hviding}, Raphael E. and {Brammer}, Gabriel and {Simcoe}, Robert A. and {Bose}, Sownak and {Bouwens}, Rychard and {Dayal}, Pratika and {Eilers}, Anna-Christina and {Fei}, Qinyue and {Furtak}, Lukas J. and {Gottumukkala}, Rashmi and {Goulding}, Andy and {Heintz}, Kasper E. and {Hirschmann}, Michaela and {Kokorev}, Vasily and {Leja}, Joel and {Liu}, Zhaoran and {Natarajan}, Priyamvada and {Santarelli}, Andrew D. and {Setton}, David J. and {Smith}, Aaron and {Tacchella}, Sandro and {Volonteri}, Marta and {Walter}, Fabian and {Weibel}, Andrea and {Williams}, Christina C.},
        title = "{Little Red Dot {\ensuremath{-}} Host Galaxy = Black Hole Star: A Gas-Enshrouded Heart at the Center of Every Little Red Dot}",
      journal = {The Open Journal of Astrophysics},
         year = 2026,
        month = may,
       volume = {9},
        pages = {62505},
          doi = {10.33232/001c.162505},
archivePrefix = {arXiv},
       eprint = {2601.20929},
 primaryClass = {astro-ph.GA},
       adsurl = {https://ui.adsabs.harvard.edu/abs/2026OJAp....962505S}
}

@ARTICLE{Begelman2026,
       author = {{Begelman}, Mitchell C. and {Dexter}, Jason},
        title = "{Little Red Dots as Late-stage Quasi-stars}",
      journal = {\apj},
         year = 2026,
        month = jan,
       volume = {996},
       number = {1},
          eid = {48},
        pages = {48},
          doi = {10.3847/1538-4357/ae274a},
archivePrefix = {arXiv},
       eprint = {2507.09085},
 primaryClass = {astro-ph.GA},
       adsurl = {https://ui.adsabs.harvard.edu/abs/2026ApJ...996...48B}
}
\bibliographystyle{aasjournal}
\end{document}